\documentclass[]{spie}  
\usepackage[]{graphicx}
\usepackage[]{aas_macros}

\newcommand{\rmicron}{$\mu$m}

\newcommand{\arcm}{$^{\prime}$}
\newcommand{\arcs}{$^{\prime\prime}$}

\newcommand{\degr}{\hbox{$^\circ$}}

\newcommand{\reg}{\textsuperscript{\textregistered}}
\newcommand{\tm}{\textsuperscript{\texttrademark}}

\title{Thermal design and performance of the balloon-borne large aperture submillimeter telescope for polarimetry BLASTPol} 

\author{J.$\,$D. Soler\supit{a,b}\supit{*}, P.$\,$A.$\,$R. Ade\supit{c}, F.$\,$E. Angil\`{e}\supit{d}, S.$\,$J. Benton\supit{e}, M.$\,$J. Devlin\supit{d}, B. Dober\supit{d}, L.$\,$M. ~Fissel\supit{b,f}, Y. Fukui\supit{g}, N. Galitzki\supit{d}, N.$\,$N. Gandilo\supit{b}, J. Klein\supit{d}, A.$\,$L. Korotkov\supit{h}, T.$\,$G. Matthews\supit{i}, L. Moncelsi\supit{j}, A. Mroczkowski\supit{k}, C.$\,$B. Netterfield\supit{b,e,l}, G. Novak\supit{i}, D. Nutter\supit{c}, E. Pascale\supit{c}, F. Poidevin\supit{m}, G. Savini\supit{n}, D. Scott\supit{o}, J.$\,$A. Shariff\supit{b}, N.$\,$E. Thomas\supit{p}, M.$\,$D. Truch\supit{d}, C.$\,$E. Tucker\supit{c}, G.$\,$S. Tucker\supit{h}, D. Ward-Thompson\supit{q}.
\skiplinehalf
\supit{a}Institute d'Astrophysique Spatiale. CNRS. France; \\
\supit{b}Department of Astronomy and Astrophysics. University of Toronto. Canada; \\
\supit{c}School of Physics and Astronomy. Cardiff University. UK; \\
\supit{d}Department of Physics. University of Pennsylvania. USA; \\
\supit{e}Department of Physics. University of Toronto. Canada; \\
\supit{f}CIERA. Northwestern University, Evanston, IL USA; \\
\supit{g}Graduate School of Science. Nagoya University. Japan; \\
\supit{h}Department of Physics. Brown University. USA; \\
\supit{i}Department of Physics \& Astronomy. Northwestern University. USA; \\
\supit{j}Department of Physics. California Institute of Technology. USA; \\
\supit{k}US Naval Research Laboratory. USA; \\
\supit{l}Canadian Institute for Advanced Research. Canada; \\
\supit{m}Instituto de Astrof\'{i}sica de Canarias. Spain; \\
\supit{n}Department of Physics. University College London. UK; \\
\supit{o}Department of Physics. \& Astronomy. University of British Columbia. Canada; \\
\supit{p}Department of Physics. University of Miami. USA; \\
\supit{q}Jeremiah Horrocks Institute of Maths, Physics, and Astronomy. UK;
}

\authorinfo{\supit{*}Corresponding author; E-mail: jsolerpu@ias.u-psud.fr, Telephone: +33 1 69 85 86 20\\
Copyright 2014 Society of Photo-Optical Instrumentation Engineers. One print or electronic copy may be made for personal use only. Systematic reproduction and distribution, duplication of any material in this paper for a fee or for commercial purposes, or modification of the content of the paper are prohibited.}
 
\begin{document} 
  \maketitle 

\begin{abstract}
We present the thermal model of the Balloon-borne Large-Aperture Submillimeter Telescope for Polarimetry (BLASTPol). This instrument was successfully flown in two circumpolar flights from McMurdo, Antarctica in 2010 and 2012. During these two flights, BLASTPol obtained unprecedented information about the magnetic field in molecular clouds through the measurement of the polarized thermal emission of interstellar dust grains. The thermal design of the experiment addresses the stability and control of the payload necessary for this kind of measurement. We describe the thermal modeling of the payload including the sun-shielding strategy. We present the in-flight thermal performance of the instrument and compare the predictions of the model with the temperatures registered during the flight. We describe the difficulties of modeling the thermal behavior of the balloon-borne platform and establish landmarks that can be used in the design of future balloon-borne instruments. 
\end{abstract}

\keywords{BLASTPol, submillimeter, polarimeter, balloon-borne telescope, thermal design}

\section{INTRODUCTION}\label{sec:intro}

The Balloon-borne Large-Aperture Submillimeter Telescope for Polarimetry (BLASTPol) is designed to map linearly polarized thermal emission from aspherical dust grains in molecular clouds \cite{galitzki2014,fissel2010,pascale2008}. BLASTPol is the first submillimeter polarimeter with sufficient mapping speed to trace polarized emission across entire molecular clouds, a critical observation to understand the role of the magnetic field in the process of star formation \cite{mckee2007,shu1987}. 

One of the main challenges of BLASTPol and other balloon-borne experiments is the effect of a changing thermal environment. The payload is exposed to air temperatures ranging from less than $-$40\degr C in the tropopause to over 40\degr C at ground level. At float altitude, the atmospheric pressure is around 5~mbar ($0.005$~atm), making convective coupling to the air practically negligible, thus the instrument has to rely on radiation as the main mechanism for cooling. The temperature control of the balloon-borne platform is made by controlling the surface area, radiative environment, and surface coating. For example: directly exposed to the sunlight and only cooling radiatively at float, a $1\,{\rm m}^2$, bare aluminum plate would reach a maximum temperature close to 200\degr C ; if the same plate is painted white, the maximum temperature is approximately 14\degr C; if the plate is painted white and has a thermal load of 10~W, the maximum temperature is 15\degr C; if the bare aluminum plate is enclosed in an aluminized mylar shield open in the anti-Sun direction, the maximum temperature is approximately 90\degr C; if the plate inside the shield is painted white, the maximum temperature is close to $-$15\degr C; and if the plate inside the shield is painted white and has a thermal load of $10\,$W, the maximum temperature is approximately $-$10\degr C.

The BLASTPol gondola is composed of over 100 conductively coupled elements. There are over 30 different heat loads produced by the power consumption of the electronic components. The level of complexity of this thermal system requires a Computer-Assisted Design (CAD) model of the gondola, which accounts for the ray-tracing of the solar and thermal radiation between elements of the gondola, something which has been done in the past using view factor coefficients specific to particular geometries. Finding the solutions of the system of linear equations resulting from the thermal equilibrium is a computationally challenging problem that has to be tackled using iterative methods. This problem is addressed in the thermal modeling of BLASTPol by using Thermal Desktop\reg, a computational tool, which integrates finite difference methods and (CAD) for modeling thermal problems.

This document describes the BLASTPol thermal model and it is organized as follows. Section \ref{Thermal:LBDenvironment} introduces the thermal environment during the balloon flight. Section \ref{Thermal:Solutions} describes the methods used to solve thermal problems in the components of the balloon-borne payload. Section \ref{Thermal:ThermalDesktopModel} describes the implementation of the thermal model in Thermal Desktop\reg. Section \ref{Thermal:BLASTPol} describes the BLASTPol thermal model and compares its predictions with the results of BLASTPol long-duration balloon (LDB) flights from Antarctica in December 2010 (BLASTPol10) and 2012 (BLASTPol12).

\section{The Balloon-borne Flight Thermal Environment}\label{Thermal:LBDenvironment}

The goal of the thermal design is to maintain all the elements of the experiment, in the absence of air convection, at temperatures below their maximum temperature ratings. Additionally, all the elements have to be kept at temperatures over their minimum temperature ratings when the air couples them to cold layers of the atmosphere. These goals are established by two separate stages of the balloon-borne flight: the ascent and the float.

\subsection{Ascent}

Heat transfer by conduction is described by Fourier's Law of Thermal Conduction \cite{incropera2007}:
\begin{equation}\label{genconductionEQ}
\frac{d\mathbf{q}}{dA} = -k \mathbf{\nabla}T\, \ \left[\mbox{W}/\mbox{m}^{2}\right],
\end{equation}
where $\mathbf{q}$ is the heat transfer rate per unit of time and normal to the element of area $A$, $k$ is the thermal conductivity, and $\mathbf{\nabla}T$ is the temperature gradient. In the 1-dimensional case, this can be approximated by
\begin{equation}\label{conductionEQ}
q = -k A \frac{dT}{dx} \, \ \left[\mbox{W}\right],
\end{equation}
where $A$ is the cross-sectional area normal to the temperature gradient. The thermal conductivity $k$ is a property of the materials. For fluids, $k$ depends on the pressure and temperature in general. However, in the case of air at densities and pressures between sea level and float altitudes, $k$ is not expected to change significantly \cite{conductivity}. 

On the surface of the Earth, the temperature of the experiment is dominated by convective air cooling. The heat transfer by convection is described by \cite{incropera2007}
\begin{equation}\label{genconvectionEQ}
q = h \Delta T\, \ \left[\mbox{W}\right],
\end{equation}
where $h$ is the convective heat transfer coefficient. The calculation of $h$ is non-trivial, since it depends on the flow properties and the shape of the object. Air at high altitude is less dense than air at sea level, reducing its convective capability and overall heat capacity. The temperature and density dependence of the Grashoff number (as defined in Ref.~[\citenum{convection}]) dictates that a component's convective coupling to the air decreases with altitude.

\begin{figure}
\centerline{\includegraphics[angle=270, width=0.46\linewidth]{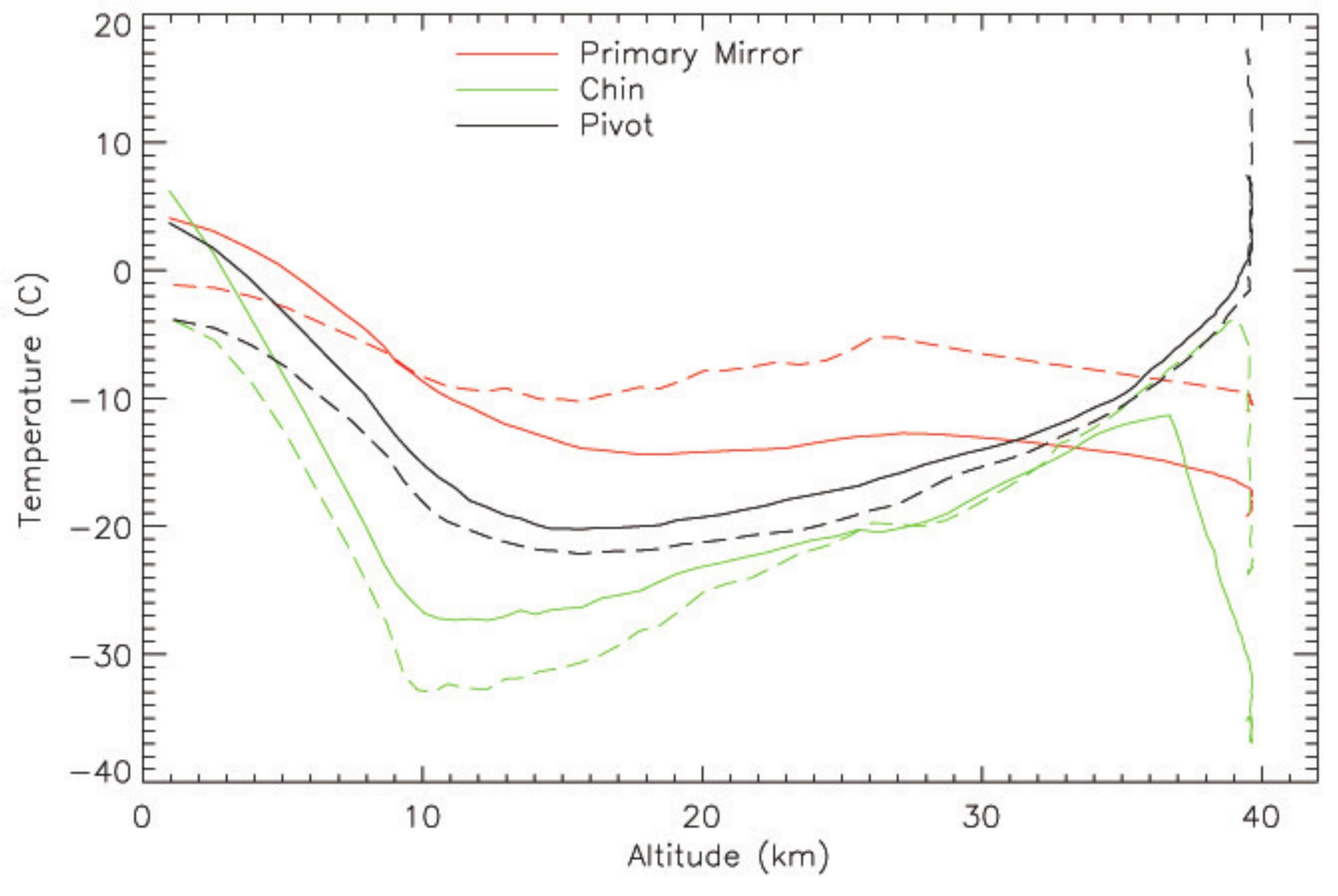}
\hspace{0.3cm}
\includegraphics[angle=270, width=0.46\linewidth]{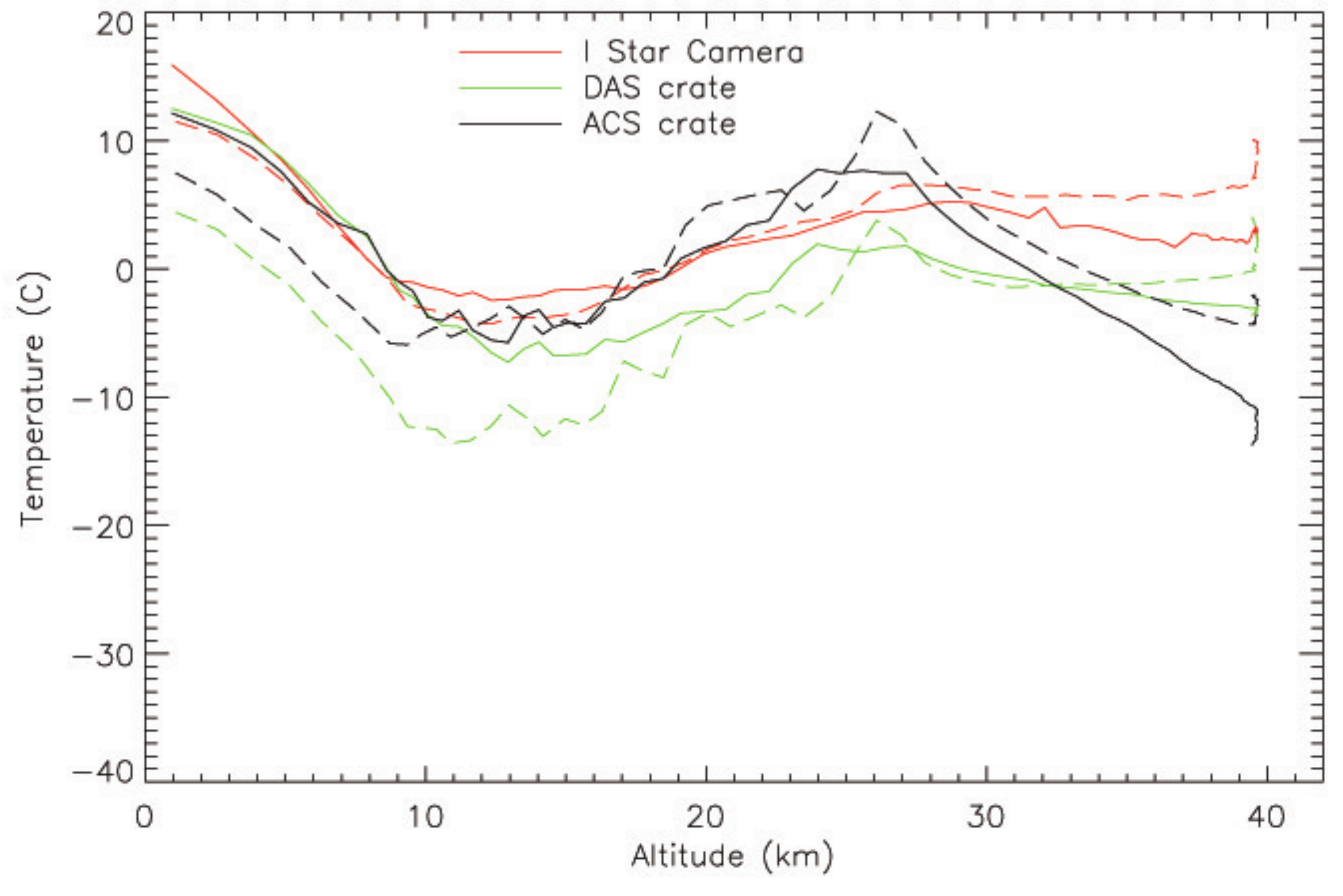}}
\caption[Temperatures of BLASTPol Gondola Elements During Ascent]{Temperatures of selected passive (left) and active elements (right) of the gondola during the ascent of BLASTPol10 (solid lines) and BLASTPol12 (segmented lines).}\label{BLASTpolAscent}
\end{figure}

The temperatures of passive elements at approximately $15\,$km altitude, shown in Figure \ref{BLASTpolAscent}, provide evidence for the coupling between the air and the gondola components in multiple layers of the atmosphere. This is important for the thermal design, since these layers include the tropopause, the transitional layer between the troposphere and the stratosphere from 8 to $18\,$km\ above sea level. In the tropopause, the rate at which the temperature changes with altitude (lapse rate) drops and the temperature gets as cold as $-$57\degr C ($-$70\degr F) \cite{saha2008}. It is in this layer of the atmosphere where most elements of the gondola may cease to function and water vapor condensates in exposed surfaces.

Figure \ref{BLASTpolAscent} shows the temperature of active (with heat load from electric power consumption) and passive (without heat load) elements of the experiment during the ascent. While transiting through the tropopause, the primary mirror cools down to $-$15\degr C, the pivot motor (unpowered) cools down to $-$20\degr C, and the ambient thermometer at the tip of the front sunshields reaches $-$35\degr C. Active elements in the inner frame, such as the star cameras and the breakout box reach temperatures down to $-$5\degr C, while the electronics on the outer frame cool down to approximately $-$15\degr C.

Thermal modeling of the convective heat transfer in the tropopause is very complex, given its dependence on the geometry and flow properties. Therefore, to guarantee the survival and recovery of the active elements after its transit through the tropopause, all electronics and motors have been cold air tested at the Columbia Scientific Balloon Facility (CSBF) in Palestine, TX. The readout and motion systems were operated in a sealed chamber filled with nitrogen gas at $-$40\degr C. All of the electronics and motors operated normally after reaching equilibrium temperatures around $-$30\degr C. Air condensation and low temperature in the tropopause are prevented by covering the electronics boxes with transparent polyethylene bags and enclosing them in $1\,$inch thick Styrofoam\tm (blue foam) boxes.

\begin{figure}
\centerline{\includegraphics[angle=270,width=0.45\linewidth]{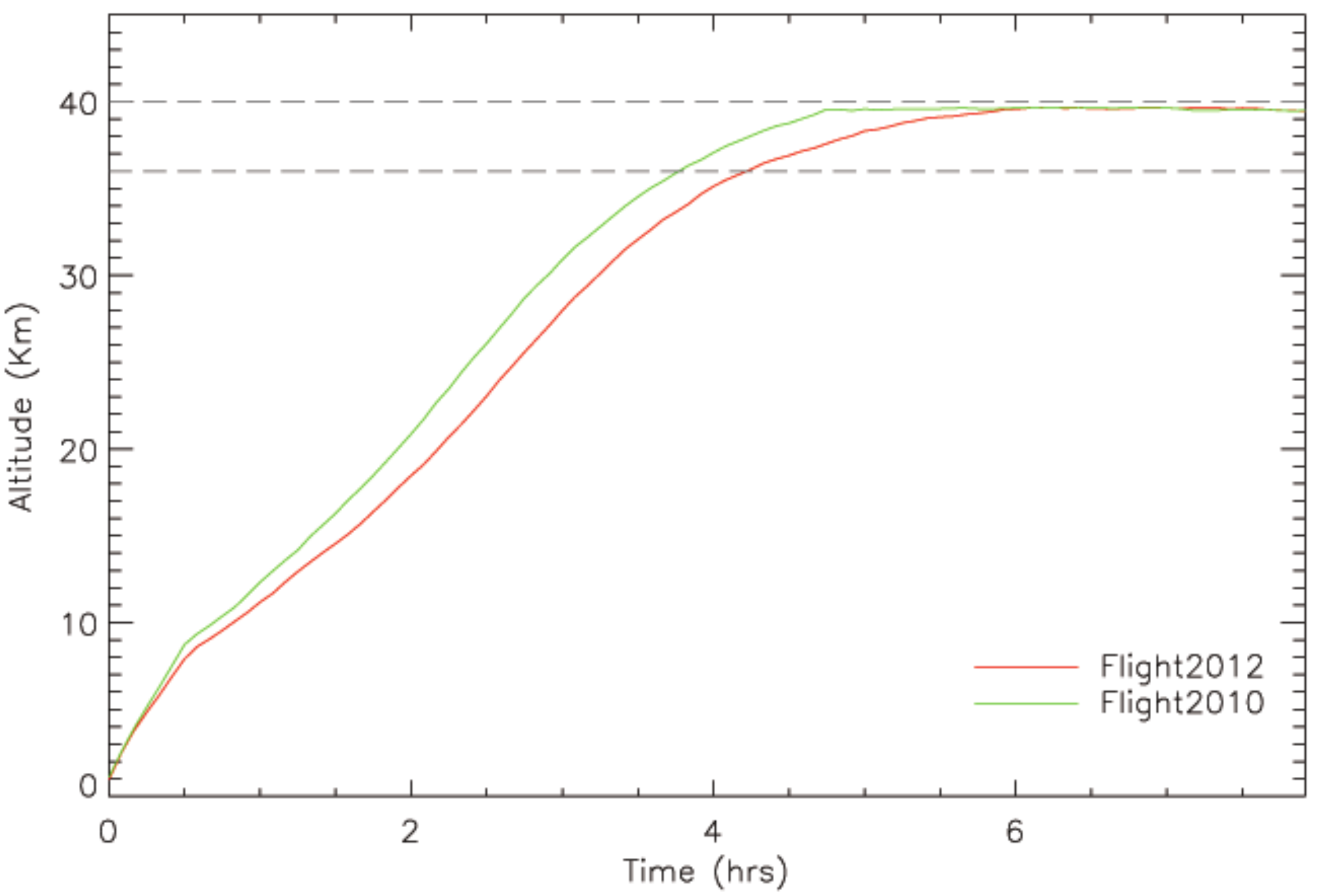}
\hspace{0.2cm}
\includegraphics[angle=270,width=0.45\linewidth]{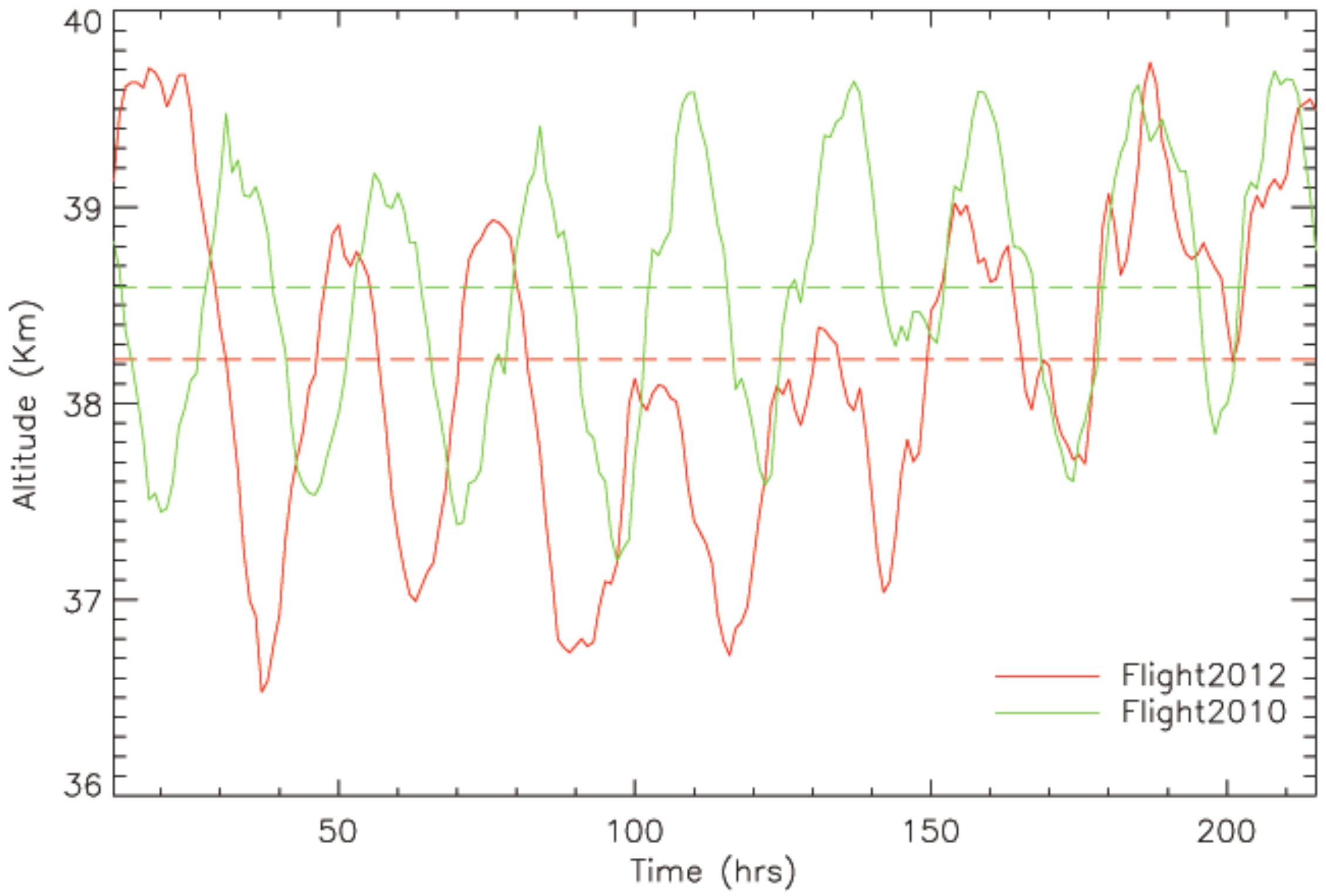}}
\vspace{0.1cm}
\caption[BLASTPol Flight Altitude]{Left: Altitude of  BLASTPol10/12 during ascent; the dashed lines indicate nominal float altitudes. Right: Altitude of BLASTPol10/12 at float; the dashed lines indicate the average elevation for each flight.}\label{BLASTpolAltitude}
\end{figure}

Figure \ref{BLASTpolAltitude} shows the altitude of BLASTPol10/12 as a function of time. The payload reaches float altitude after $\sim4\,$hours, spending a little more than $1\,$hour in the tropopause. Once at float, the altitude of the payload changes diurnally between 40 and $37\,$km. This change in altitude is caused by the expansion and contraction of the helium in the balloon produced by diurnal cycles.

\subsection{Thermal Environment at Float}

The right panel of Figure \ref{BLASTpolAltitude} shows the float altitude and its change with the diurnal cycles in BLASTPol10/12. At a mean float altitude higher than $\sim38\,$km, the telescope is above 99\% of the atmosphere, which is 98\% transparent across the BLASTPol bands \cite{wiebe_thesis2008}. At this altitude, the atmospheric pressure is $\sim5\,$mbar ($0.005\,$atm) and the air density is less than 7.3$\times$10$^{-4}\,$kg$\,$m$^{-3}$. In this near to vacuum environment, little or no thermal convection should take place, except at surface boundary layers. The temperatures of the gondola components are determined by radiation emission and absorption, and by conductive coupling between elements and from these elements to the air.

The power transfer by radiation is described by the Stefan-Boltzmann law \cite{incropera2007}:
\begin{equation}\label{Radiation}
P^{(\mbox{rad})} = \epsilon A \sigma T^{4}\, \ \left[\mbox{W}\right],
\end{equation}
where $\sigma=5.67\times 10^{-8}\,$W$\,$m$^{-2}\,$K$^{-4}$ is the Stefan-Boltzmann constant. $A$ and $T$ are the effective surface area and temperature of the object under consideration. The emissivity, $\epsilon$, is the fraction of the thermal radiation that is emitted. The thermal equilibrium for a particular element of the gondola is described by
\begin{equation}\label{RadiationEquilibrium}
 P + \sum \alpha_{i} A_{i} I_{i} = \epsilon A \sigma T^{4}.
\end{equation}
Here $P$ is the heat load resulting from the electrical power consumed by the element, $I_{i}$ is the intensity of each external radiation source illuminating the element, integrated over the relevant wavelength band, $A_{i}$ is the effective surface area as seen from each radiation source, and $\alpha_{i}$ is the percentage of the incoming radiation that is absorbed by the element.

The emissivity $\epsilon$ is wavelength dependent, but for equilibrium temperatures around $300\,$K the dominant thermal emission is produced around $10\,$\rmicron. The effective surface area $A_{i}$ depends on the geometry of the gondola and the relative orientation of the elements on it; this is usually not trivial to calculate by hand and it is estimated using  CAD thermal modeling and ray tracing algorithms. The absorptivity $\alpha$ is also wavelength dependent. However, the dominant absorptivity considered in the study of the gondola is that around $500\,$nm, the central frequency of sunlight, since most of the illumination comes from direct or reflected sunlight.

A passive element with equal radiative and air-conducting area, $A$, painted white, placed in a radiative environment at $300\,$K, would reach equilibrium temperatures of $298.8\,$K in the absence of air coupling and 263.62 and 246.24~K if coupled to the air at 270 and $250\,$K, respectively. If this element is the Attitude Control System (ACS) crate of BLASTPol, with a heat load of $\sim27\,$W and $A\sim0.1\,$m$^{2}$, its equilibrium temperature would be 298.8~K in the absence of air conduction and 283.61 and 272.4~K if coupled to air at 270 and $250\,$K, respectively. 

Given the difficulty in determining the air temperature at float altitudes and separating its effect from the changing radiative environment, the BLASTPol thermal model does not rely on the conductive coupling to the air. The net effect of conductive coupling to the atmosphere would be a modulation of the temperatures, bringing them closer to the air temperature. However, we chose to use a worst case scenario where the temperature ranges coming from the radiative heat transfer and conduction between gondola elements in the absence of air. This approach successfully maintained the integrity of the experiment in BLASTPol10/12, as discussed in Section \ref{Thermal:BLASTPol}.

\begin{figure}
\centerline{\includegraphics[width=0.45\linewidth]{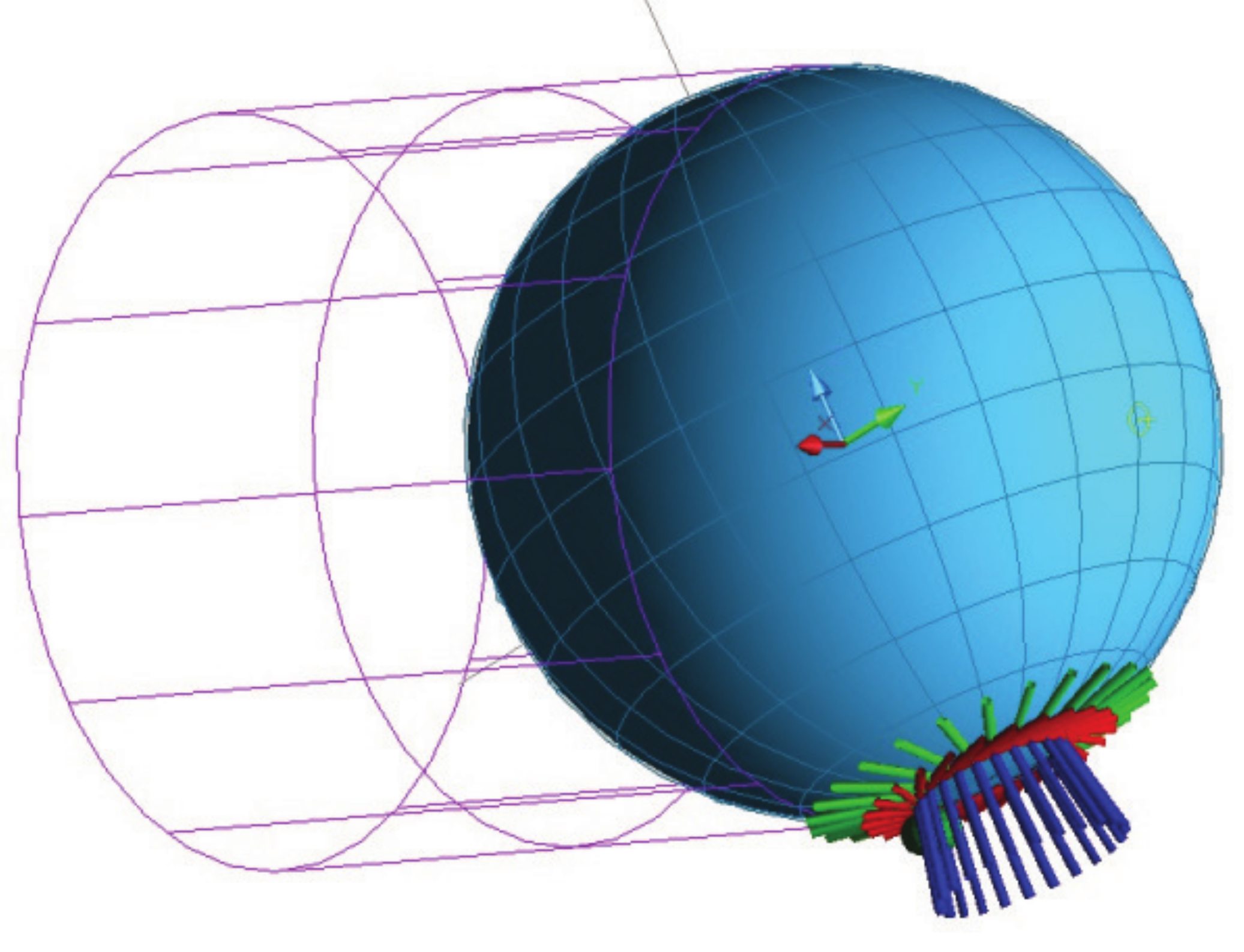}
\hspace{0.5cm}
\includegraphics[width=0.32\linewidth]{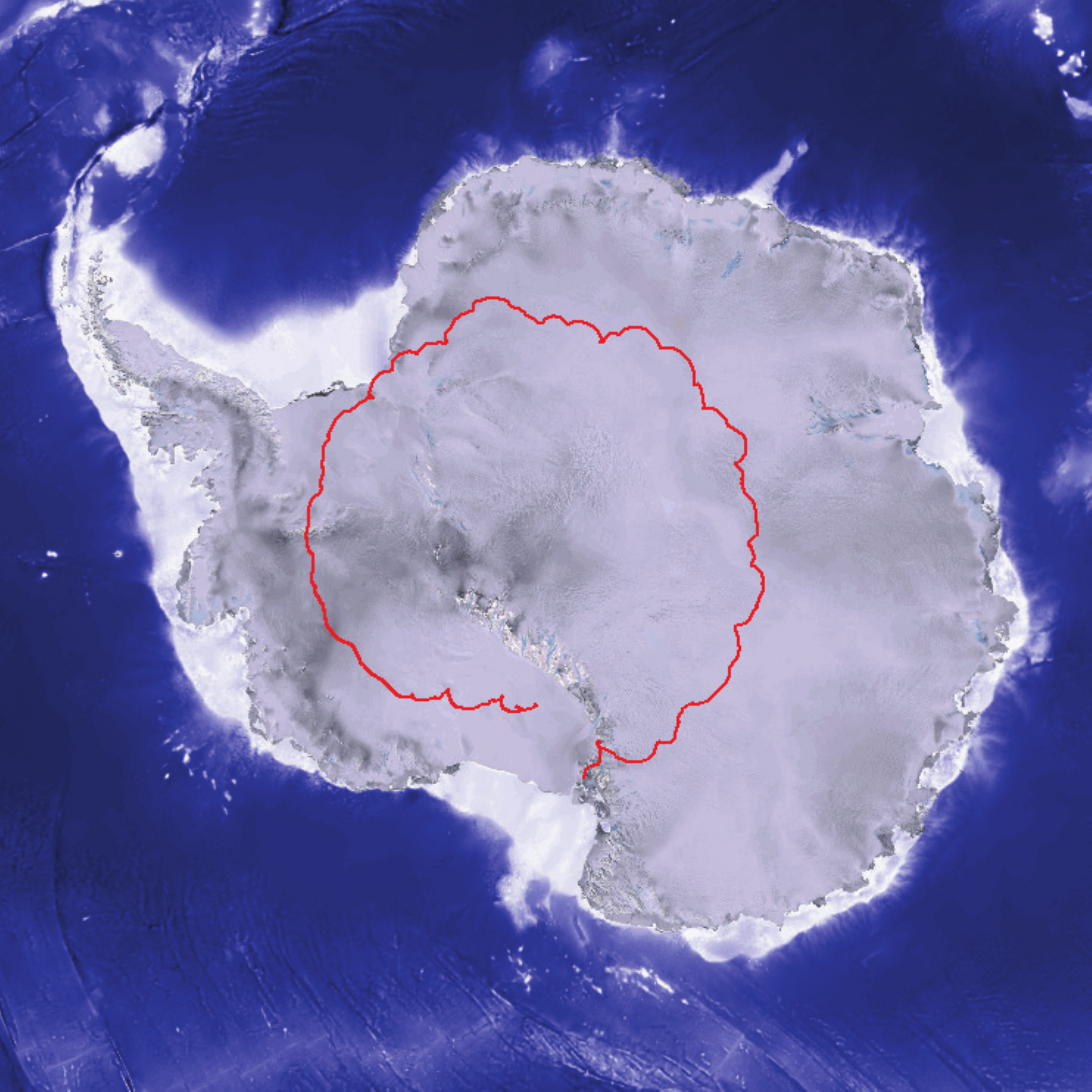}}
\caption[Model of the Antarctic Flight Trajectory]{Left: Rendering of the orbit defined for the BLASTPol thermal analysis. This is a circumpolar orbit at latitude 78\degr\ South. The cylinder corresponds to the projected shadow of the Earth, with the Sun as a backlight source. Right: BLASTPol2012 circumpolar flight trajectory. The image shows the ground optical properties relevant in the estimation of the albedo: snow (gray); ice shelf (white); and sea water (dark blue).}\label{BLASTpolThermalOrbit}
\end{figure}

External illumination of the payload comes from three main sources: solar radiation; reflected sunlight from the surface of the Earth (albedo); and the thermal emission from the surface of the Earth (planet shine). The contribution of each one of these sources depends on the trajectory of the flight and the time of year. The model of the thermal environment at float considers the Earth as a sphere with a $6356.8\,$km radius, axial tilt of 23\degr26\arcm24\arcs.41, located 149.60$\times10^{6}\,$km from the Sun, and solar illumination as discussed below. The payload is simulated in a circumpolar orbit at fixed latitude 78\degr\ South at $38\,$km above sea level, starting on December 22, as shown in Figure \ref{BLASTpolThermalOrbit}.

\subsubsection{Solar Radiation}

The spectrum of the solar radiation is close to that of a blackbody with a temperature of about 5,800$\,$K. The sunlight flux density at the top of Earth's atmosphere is 1,366$\,$W$\,$m$^{-2}$. It is composed, dividing by total energy, of about 50\% infrared light ($\lambda > 0.7\,$\rmicron), 40\% visible light ($0.7\,$\rmicron$\,<\lambda < 0.4\,$\rmicron), and 10\% ultraviolet light ($\lambda < 0.4$~\rmicron) \cite{holton2003encyclopedia}.

\subsubsection{Albedo}

In the context of the BLASTPol thermal model, the albedo is defined as the ratio of reflected radiation from the surface of the Earth to incident radiation upon it. Being a dimensionless fraction, it may also be expressed as a percentage, and is measured on a scale from zero (for no reflected power for a perfectly black surface), to unity (for perfect reflection from a white surface). In detailed thermal modeling, a bidirectional reflectance distribution function may be required to characterize the scattering properties of a surface accurately, although for the purpose of the thermal modeling of the payload, the albedo is a sufficient first approximation.

The average overall albedo of Earth (planetary albedo) is 30 to 35\%, because of cloud cover, but locally varies widely across the surface, depending on geological and environmental features. In Antarctica, the surface of the continent is covered by snow, which can be as much 95\% reflective, while the surface of the Antarctic sea reflects only 10 to 15\% of the incoming sunlight. The historical record of Antarctic flights indicates that drift to latitudes outside of the continent is unlikely; therefore an albedo of 95\% is used in the BLASTPol thermal model. This was found to be an accurate estimation of the flight conditions in 2012, as shown in the right panel of Figure \ref{BLASTpolThermalOrbit}.

\subsubsection{Planetary Radiation}

The planetary radiation, also called \emph{planetshine}, is the thermal emission by the surface of the Earth. The temperature of the surface of the Earth is $\sim$270\degr K, therefore the \emph{planetshine} can be modeled as a blackbody peaking at $10\,$\rmicron. The radiance integrated between 8~\rmicron\ and 15~\rmicron\ is $38\,$W$\,$m$^{-2}\,$sr$^{-1}$, which at $38\,$km from sea level corresponds to a flux density $237\,$W$\,$m$^{-2}$. In that frequency range, the emissivity of the ground is close to 0.99 \cite{modislib}. The flux density from \emph{planetshine} reaching the balloon-borne platform is around 17\% of the solar flux density.

An additional source of illumination is the diffuse sky radiation, which is solar radiation scattered in elastic processes such as Rayleigh or Mie scattering by molecules in the atmosphere\cite{bohren2007}. Diffuse sky radiation can also be infrared when the scattering processes are inelastic and the light is re-emitted with a longer wavelength. Although diffuse sky radiation can be relevant for the optical performance a balloon-borne telescope, it does not significantly contribute to the thermal performance of the payload and therefore is not considered in the thermal model.

\section{Thermal Solutions}\label{Thermal:Solutions}

Given the thermal conditions of the balloon-borne flight described in Section \ref{Thermal:LBDenvironment}, strategies for the control of the thermal performance of the experiment can be: shielding of the elements from direct exposure to sunlight; selection of surface coatings to enhance radiative cooling; and finally, conductive coupling to radiators.

\subsection{Shielding}

BLASTPol flies during the Antarctic summer, which means that the payload is continuously exposed to sunlight during the whole flight. Extensive shielding is needed to regulate the temperature of the instrument. The sunshields define the azimuth ranges relative to the Sun, where direct sunlight and albedo are blocked from the optics and all electronics, except for the pivot and the Sun sensor. The azimuth range defines the portions of the sky that can be observed by the telescope. The sunshield geometry determines the targets and sky coverage of the experiment.

BLASTPol is surrounded by shields comprised of Duralar\tm $0.002\,$inch thick Mylar\reg\ or Polyethylene Terephthalate (PET) film, with a 2$\times10^{-3}\,$inch layer of Vapor-Deposited Aluminum (VDA) and A-25-NPS Lamart 3$\times10^{-3}\,$ Mylar\reg\ film, bonded to 3.5$\times10^{-4}\,$inch aluminum foil. The vapor-deposited film was mounted to the outer face of the shields, while the bonded film is mounted to the inner face, with the non-conducting (mylar) side facing out on all surfaces that may be exposed to the Sun. The VDA film has a greater emissivity, which optimizes radiative cooling to the outside of the gondola. The bonded film shows low in-band transmission, such that no Styrofoam\tm layer is needed. This configuration showed no considerable ultraviolet (UV) degradation during BLASTPol10/12. Both mylar layers are taped directly to the white-painted aluminum sunshield frames.

\begin{table}[htbp!]
\caption{Optical properties of the surface coatings used in the BLASTPol thermal model\cite{kauder2005}.}\label{OpticalProperties}
\begin{center} 
  \begin{tabular}{|l|c|c|c|c|}
    \hline
    Material & Solar  & Infrared & $\alpha/\epsilon$                \\
             & Absorptivity ($\alpha$) & Emissivity ($\epsilon$) &  \\ \hline
    Aluminum (machined, non-polished)          & 0.080 & 0.020 & 4.000 \\
    Stainless Steel (machined, non-polished)    & 0.390 & 0.110 & 3.545 \\ \hline
    Duralar Al Mylar\reg\ (2.0~mil, VDA)         & 0.170 & 0.760 & 0.223 \\
    Lamart Al Mylar\reg\ (3.0~mil, bonded)       & 0.230 & 0.720 & 0.319 \\
    Polyethylene, LDB Balloon Film              & 0.016 & 0.050 & 0.320 \\
    Teflon Silver Tape, (5.0~mil)               & 0.080 & 0.810 & 0.098 \\ \hline
    SunCat G10/FR4                              & 0.123 & 0.940 & 0.130 \\
    SunCat Laminate Back                        & 0.630 & 0.950 & 0.663 \\ \hline
    SunCat Solar Cell                           & 0.895 & 0.940 & 0.952 \\
    White Paint                                 & 0.230 & 0.800 & 0.287 \\
    Zinc Oxide White Paint                      & 0.160 & 0.930 & 0.172 \\
    \hline
\end{tabular}
\end{center}
\end{table}

 \subsection{Conductive Coupling}

The thermal control of the active elements of the gondola can be achieved by improving their conductive coupling to large elements, which act as heat sinks, or large surface area elements, which serve as radiators. The heat transfer by conduction is described by  Equation \ref{conductionEQ}, which can be rewritten as 

\begin{equation}\label{MYconductionEQ}
q = -k \left(\frac{A}{x}\right)\Delta T \, \ \left[\mbox{W}\right].
\end{equation}
This assumes that the temperature gradient in the plane normal to the direction of the heat conduction is negligible. To first order, this is a good approximation to the heat transfer between elements of the gondola, and allows us to characterize the geometric conductive coupling by $(A/x)$, the ratio of the cross-sectional area to the distance between isothermal nodes. Throughout this document, the conductive couplings are expressed in terms of $(A/x)$. The thermo-physical properties of the materials in the BLASTPol model are taken from Ref. [\citenum{perry2007}].

The thermal model calculates the temperature of the electronics boxes, assuming that the electronic components and the box shell are in thermal equilibrium. The conductive couplings are considered to be close to ideal by reducing the gaps between interfaces using DOW Corning\reg\ silicone grease TC-5026 or Stockwell conductive silicone sponge R-10400. In some particular cases detailed modeling of internal components is necessary. For example, the model of the BLASTPol computer box includes the individual chips and couplings to the faces of the box.

\subsection{Surface Coatings}

The temperatures of the gondola elements at float are regulated changing the optical properties of their surfaces. This can be achieved by applying surface coatings. Surface coatings are characterized by the ratio of the absorbed solar radiation to emitted infrared (IR) radiation or solar absorptivity to IR emissivity $\alpha/\epsilon$. Surface coatings with $\alpha/\epsilon$ greater than 1.0, such as bare aluminum and steel, absorb more radiation than they emit, and this results in heating of the gondola elements. Surface coatings with $\alpha/\epsilon$ less than 1.0, such as white paint and mylar, emit more radiation than they absorb and this in consequence results in cooling of the gondola elements. The properties of the most common surface coatings used on the BLASTPol gondola are summarized in Table~\ref{OpticalProperties}.

The absorption and emission also depends on the effective surface area of the gondola elements. Small elements, such as electronics boxes, are particularly critical, since their effective surface area can be greatly affected by low quality painting. Additionally, the presence of multiple connectors makes it easy to underestimate the effective surface area over which the box can emit thermal radiation. Rust-Oleum\reg\ white specialty appliance epoxy (refrigerator paint) proved to be an effective solution in BLASTPol10/12. More critical elements, such as the pivot controller box, required more effective reflective and emissive coating, such as silver Teflon tape.

The thermal model allows the definition of optically inactive regions in which the radiative heat transfer is negligible. Surfaces such as the inside of the sunshield frame, the inside of the inner frame, and the inside of some boxes are defined as optically inactive to improve the speed of the simulation.

\section{Computer Assisted Thermal Modeling}\label{Thermal:ThermalDesktopModel}
The thermal modeling of BLASTPol was carried out using the thermal network analyzer SINDA/FLUINT\reg\ and the CAD based thermal analysis and design software Thermal Desktop\reg\ \cite{SindaFluint2012,ThermalDesktop2012}.
Thermal Desktop\reg\ is a program that receives as input the 3D AutoCAD\reg\ model of the experiment. Each of the elements in the model is represented by a series of surface isothermal nodes. An isothermal node is the minimum element of the thermal model. Each surface can contain one or more isothermal nodes, which define the points where the temperature is registered at the end of the calculation. 

Thermal Desktop\reg\ creates a thermal network for the conductive couplings and the geometry of the gondola. The thermal network is the input for RadCAD\reg\ a module that calculates radiation exchange factors and orbital heating rates defined by the coordinates of the flight trajectory and the illumination sources described in Section \ref{Thermal:LBDenvironment}. RadCad\reg\ solves the heat transfer case generated as a result of the equilibrium state and transient temperatures of the nodes defined in the model. The output of Thermal Desktop\reg\ and RadCAD\reg\ is combined to create inputs for SINDA/FLUINT, a thermal/fluid analyzer that iteratively solves the heat transfer equation system and generates transient or equilibrium temperatures for each node in the model.

Figure \ref{BLASTpolThermalRender} shows the 3D AutoCAD\reg\ model of the BLASTPol gondola. Each of the surfaces is modeled by multiple surface elements, and the heat transfer on them is solved by finite differences methods. Each surface can contain one or more isothermal nodes on which the output temperatures are calculated. Sensitive surfaces, such as the BLASTPol primary mirror, are segmented in multiple nodes, while less sensitive components, such as the electronics boxes, are modeled with multiple surfaces, but only one isothermal node. The thermal model of BLASTPol has over 600 surfaces and around 1,000 isothermal nodes.

The objective of the calculations performed using SINDA/FLUINT\reg\ and Thermal Desktop\reg\ is to establish the temperature ranges of the elements of the experiment at float, in order to evaluate solutions that allow their normal operation during the flight. Each model is developed using the following steps.

\begin{enumerate}
  \item Sun-shield design: An initial 3D model of the experiment is created, and depending on the science targets of each experiment, the Sun avoidance angles are defined and the geometry of the sunshields is determined.
  \item Definition of thermal ranges: A list of the thermal operational ranges of all the elements of the gondola is made. Vacuum and cold operation testing provide additional validation and redefinition of the temperature ranges.
  \item Initial Calibration: The model is run using an initial guess of the optical properties and conductive couplings, and including the heat loads defined by the electrical power consumed by each element. The results of this \emph{ansatz model} are calibrated with the temperatures registered in comparable elements of previous experiments, in this case BLAST\cite{pascale2008} and Boomerang\cite{piacentini2002,crill2003}. The main objective of this step is to identify critical elements that are expected to reach temperatures outside of their operational ranges. To facilitate the calculations the evaluation of maximum and minimum temperatures can be made identifying a \emph{cold case} and a \emph{hot case}, instead of comparing transients. The \emph{hot case} corresponds to maximum power dissipation and maximum Sun illumination; the \emph{cold case} corresponds to no power dissipation and minimum Sun illumination.
  \item Thermal solutions are implemented where needed. The thermal control strategies are described in Section \ref{Thermal:Solutions}. Elements that overheat can be coupled to elements with a larger emissive surface area that serve as radiators. Elements that run too cold can be decoupled from heat sinks or their surface can be covered with less emissive coatings. Boxes can overheat if they are surrounded by elements that obstruct radiation, so changing their location can also help.
  \item Reevaluation of the model: Once the thermal solutions are implemented the model is run again to check the effectiveness of the thermal control strategy.
  \item Evaluation of transient solutions: Once the \emph{hot case} and \emph{cold case} are evaluated, transient solutions are generated. These cover at least a couple of days during the flight and enables the evaluation of the long-term behavior of the elements in the experiment. The effect of latitude drift during the flight is important, since it results in changes in the maximum solar elevation during the day and different illumination of the gondola.
  \item Keep model updated: The thermal model is useful only if it includes the updated information of the experiment. The heat loads are updated by keeping track of the operation currents on the different elements. Any changes in the geometry or the location of the elements in the gondola are tracked.
\end{enumerate}

\section{The BLASTPol Thermal Model}\label{Thermal:BLASTPol}

A detailed description of the BLASTPol gondola components can be found in Refs.~[\citenum{galitzki2014}], [\citenum{fissel2010}], and [\citenum{pascale2008}]. The BLASTPol thermal model includes all the optical systems outside of the cryostat, the inner frame, outer frame, electronics boxes, reaction wheel, aluminum honeycomb decks, sunshield panels and frames, solar array, CSBF solar array, Support Instrumentation Package (SIP), suspension cables, pivot, ballast hopper, and the balloon. The telescope observation ranges and the thermal performance of the electronics, which evolved from the BLAST campaigns, were the main driver of the thermal design of BLASTPol.

\subsection{Sunshields}\label{thermal:blastpolsunshields}

\begin{figure}
\centerline{\includegraphics[width=0.4\linewidth]{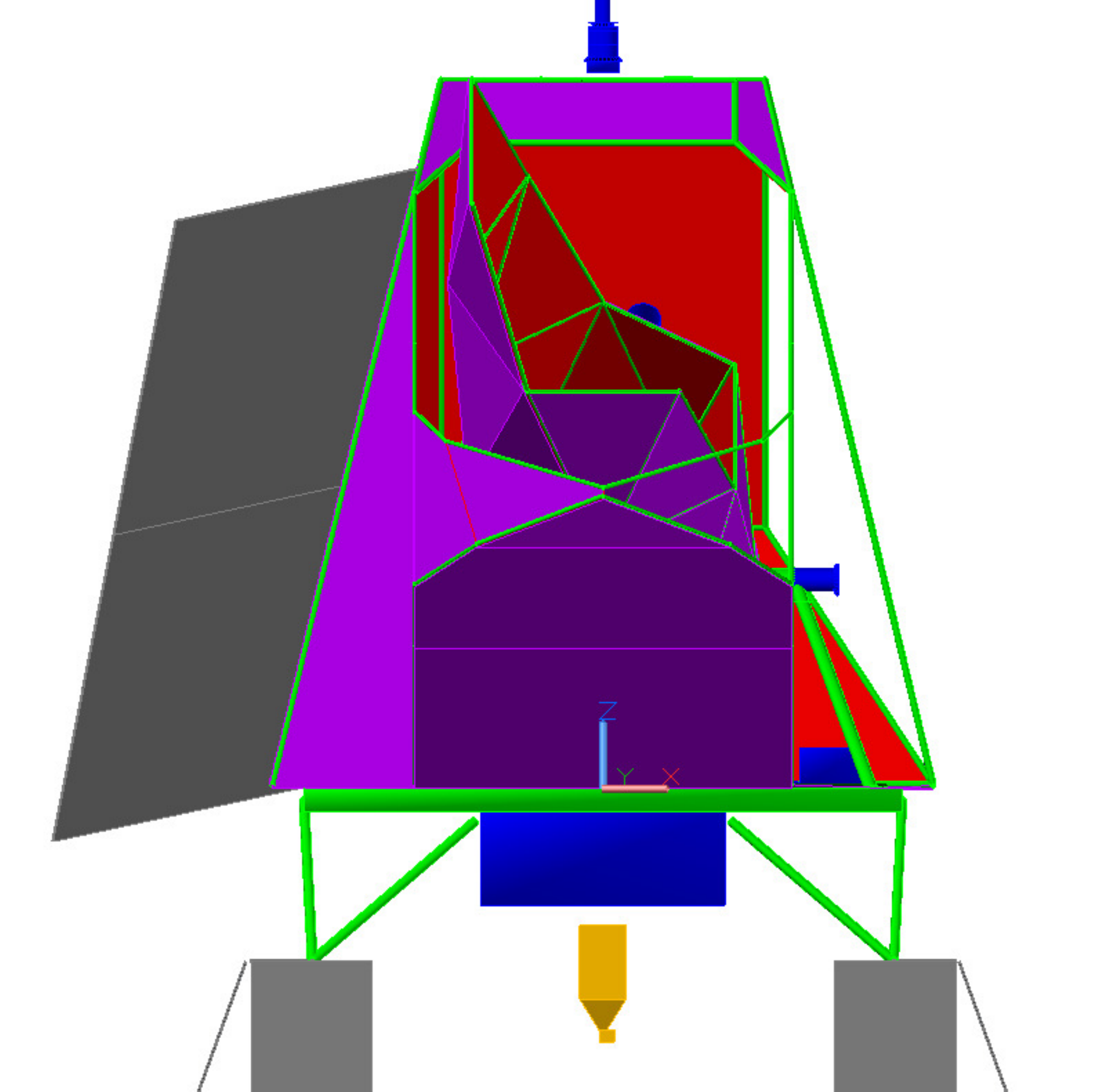}
    \includegraphics[width=0.4\linewidth]{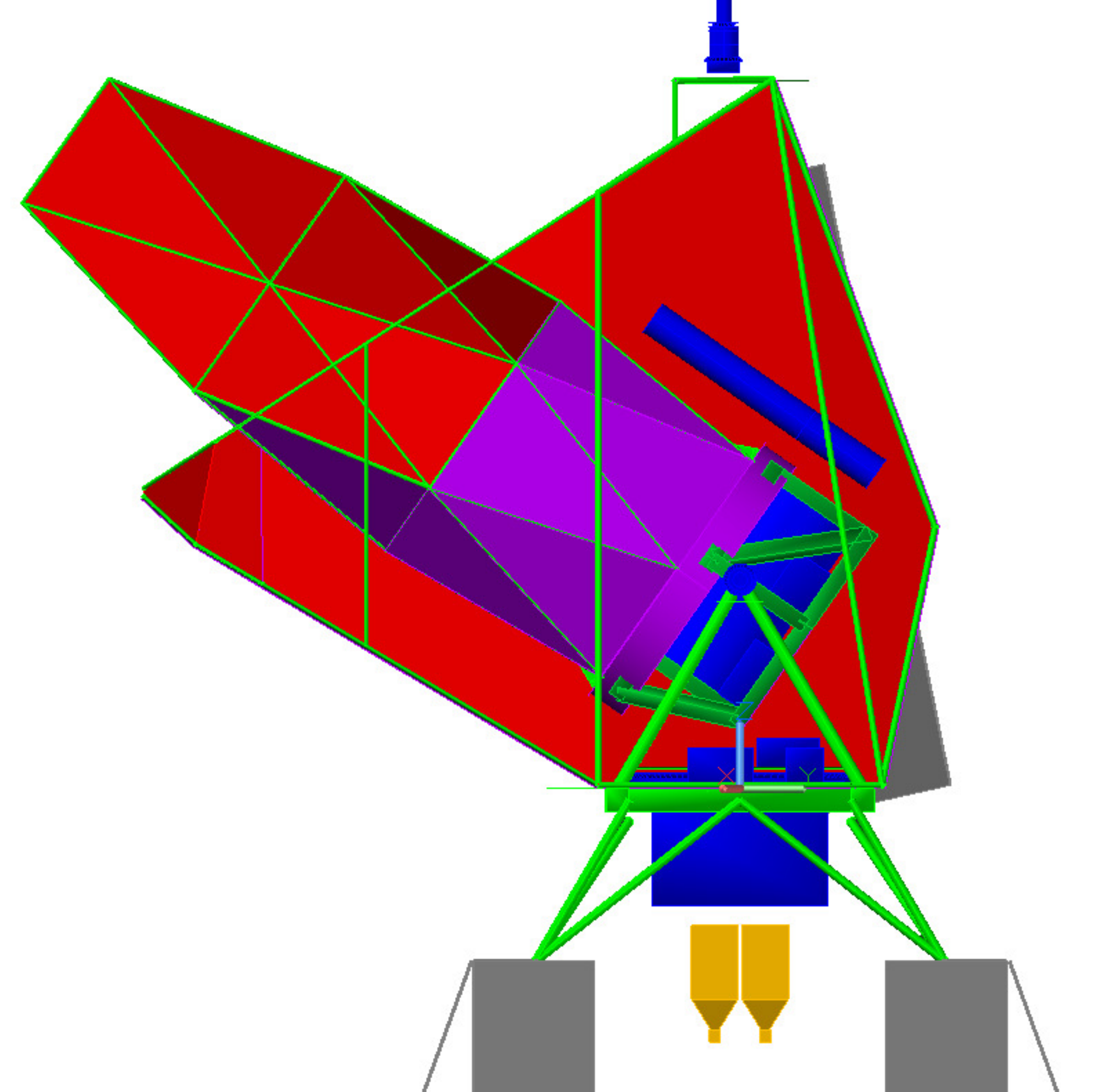}}
  \caption{Front and port side views of the BLASTPol Thermal Model. The red surfaces correspond to the inner face of the sunshields. Purple corresponds to the outer face of the sunshields. The solar arrays are shown in grey.}\label{BLASTpolThermalRender}
\end{figure}

The primary scientific goal of BLASTPol is the measurement of polarized dust emission in molecular clouds. The main targets observed from Antarctica during the Austral summer are the Vela Molecular Ridge and the Lupus Molecular Cloud. At this time of the year, Vela is located at 140\degr\ left of the Sun in azimuth and between 30\degr\ and 50\degr\ in elevation. Sun shielding for this almost anti-Sun observation is straight-forward, involving mainly covering the back of the gondola. Less straightforward is the observation of Lupus, which is located at $\sim$40\degr\ left of the Sun in azimuth and between 25\degr\ and 50\degr\ in elevation. The observation of this target required unprecedented sunshielding, which allows proximity to the direct illumination from the Sun, while maintaining the thermal stability of the experiment.

\begin{figure}
\centerline{
\vspace{-0.3cm}
\includegraphics[height=0.22\textheight]{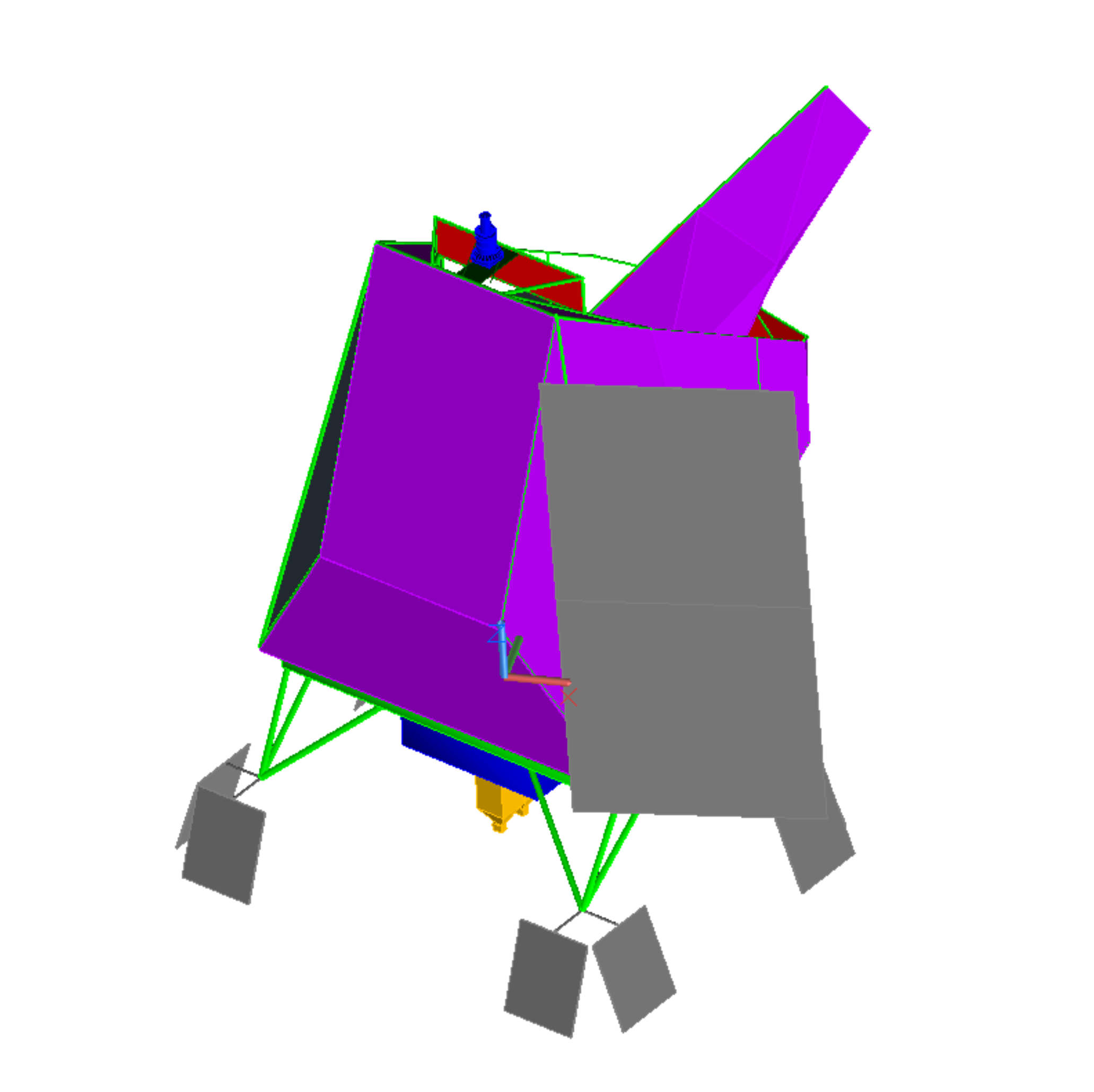}
\hspace{-0.7cm}   
\includegraphics[height=0.22\textheight]{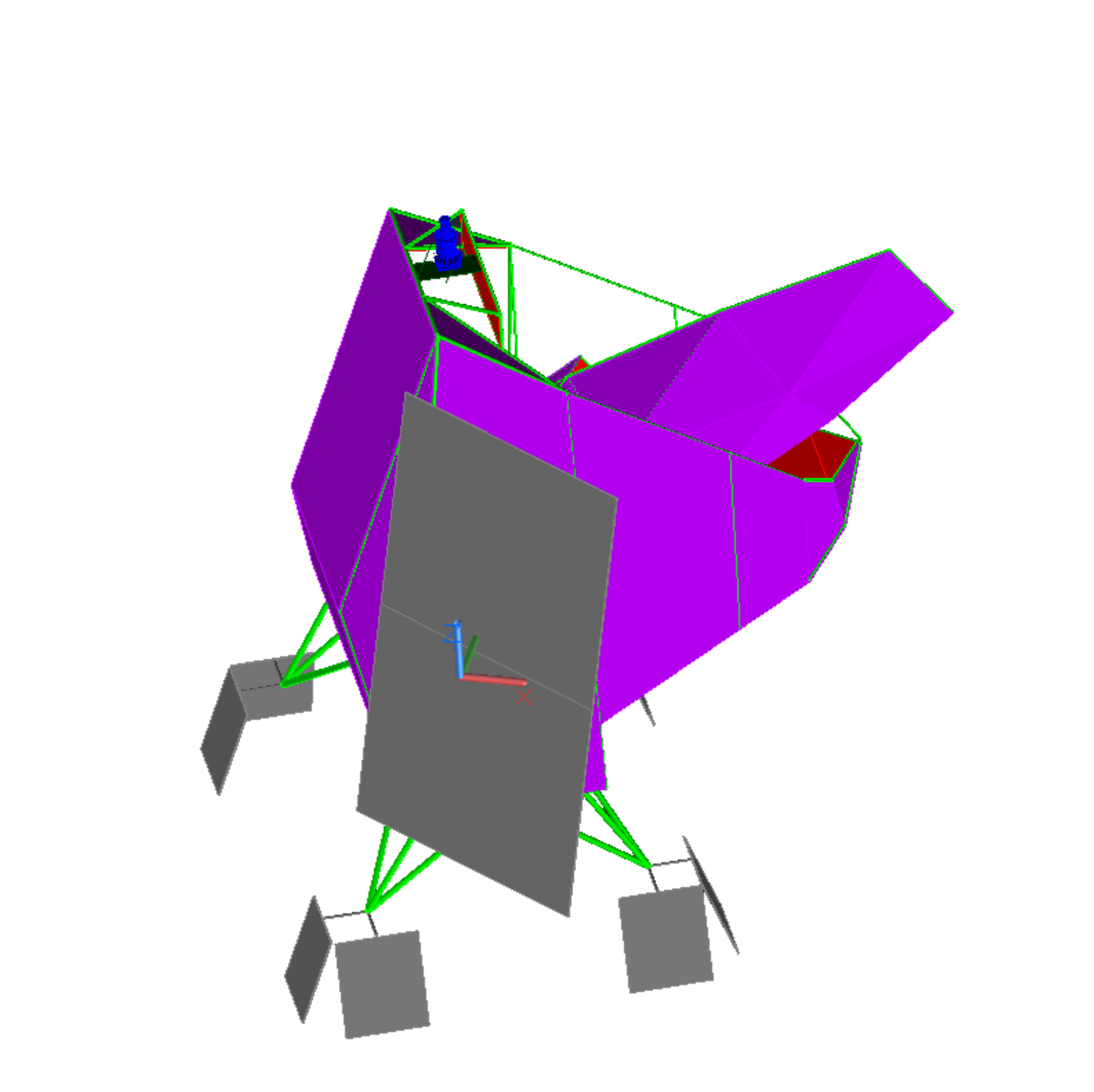}
\hspace{-0.7cm}
\includegraphics[height=0.22\textheight]{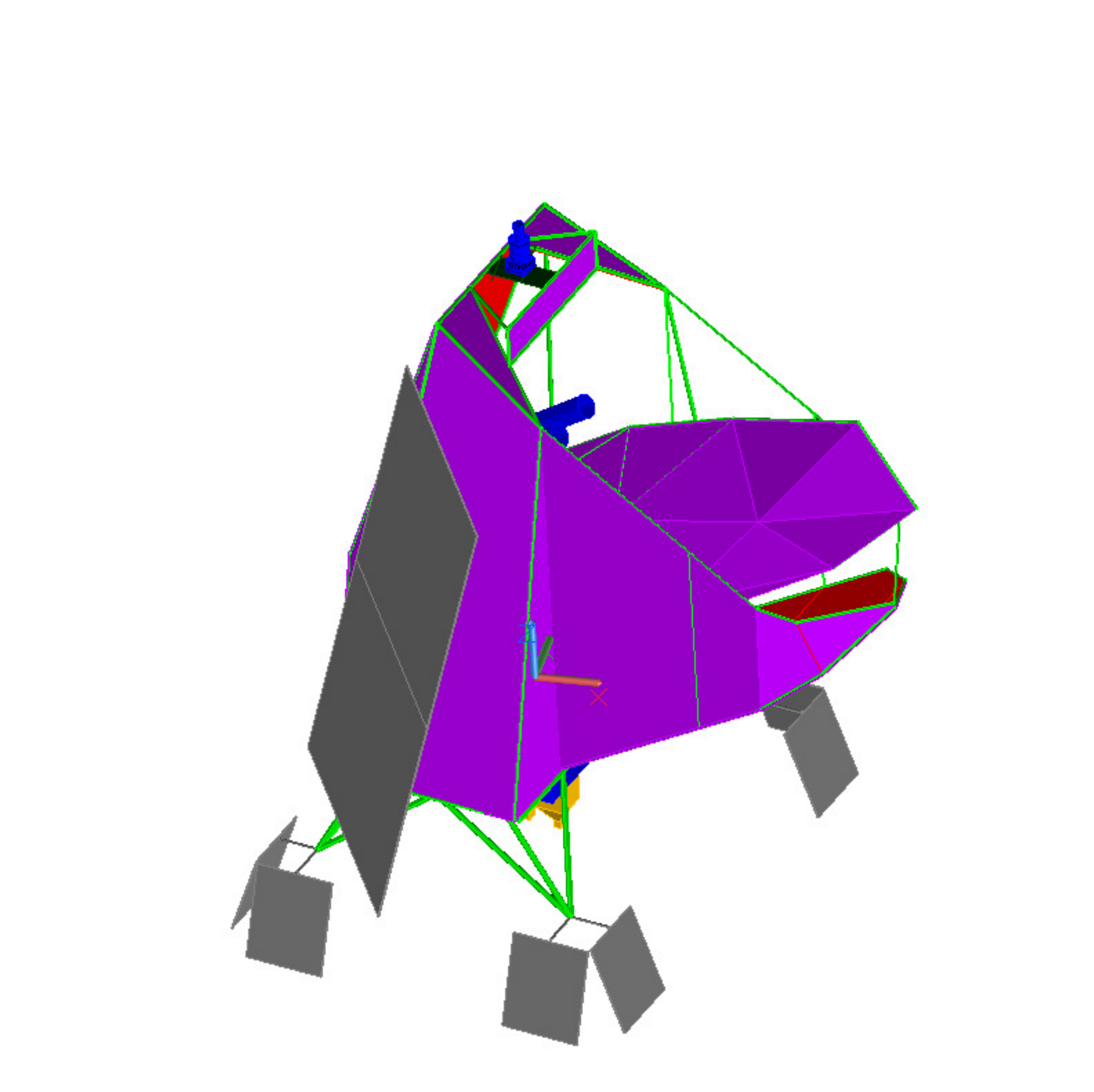}
\hspace{-0.7cm}
\includegraphics[height=0.22\textheight]{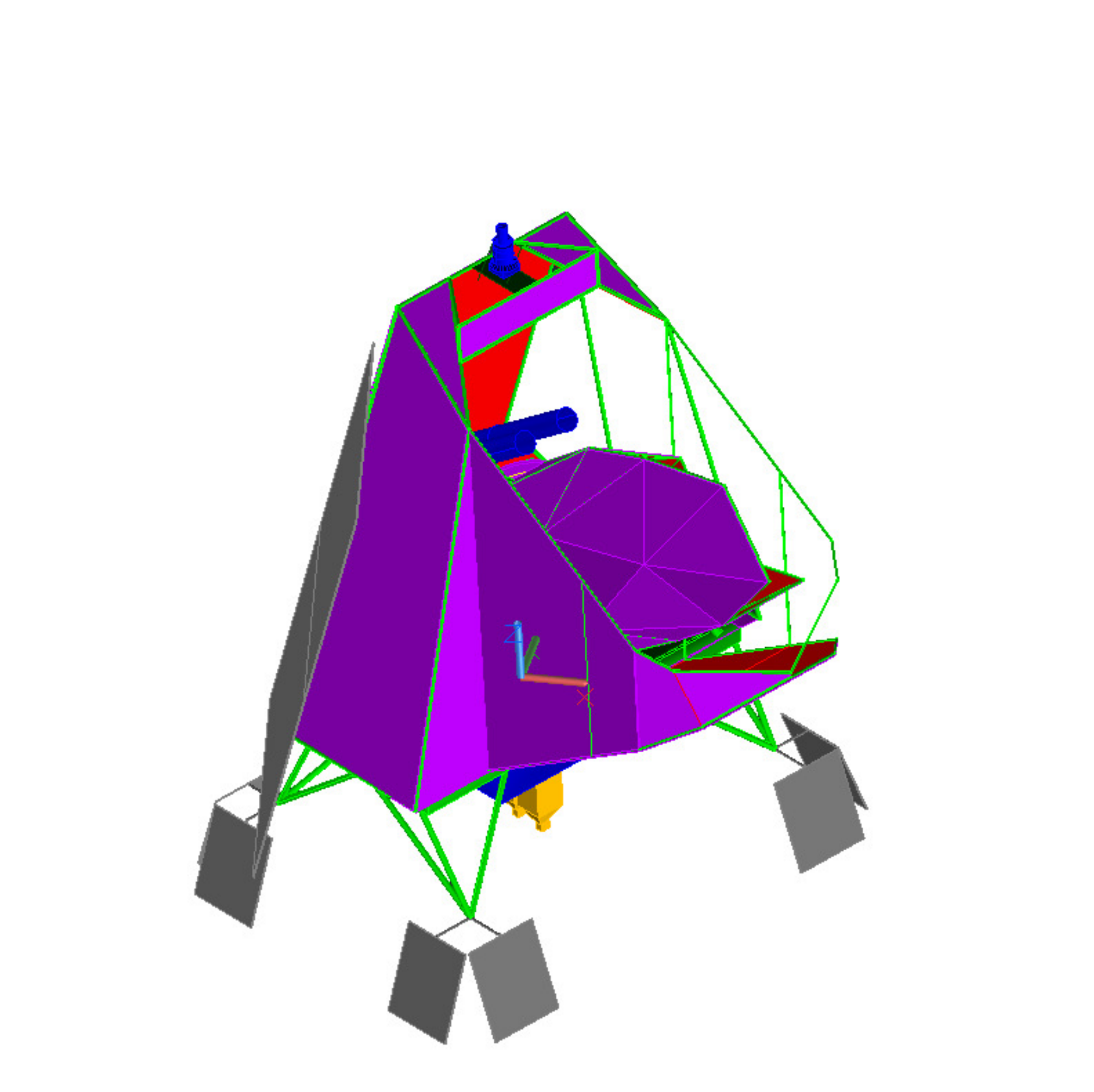}}
\vspace{0.1cm}
\caption{Views of the BLASTPol gondola from the Sun at maximum elevation at McMurdo (Lat.~77\degr\ South, Lon.~165\degr\ East). From left to right, the renderings correspond to boresight azimuth positions at 140\degr, 100\degr, 60\degr, and 40\degr\ from the Sun.}\label{BLASTpol360}
\end{figure}

The solution for the azimuth Sun-avoidance required by BLASTPol is illustrated in Figure \ref{BLASTpolThermalRender}. The outer frame of the sunshields maintains the original BLAST design described in Ref.~[\citenum{pascale2008}], although only the starboard side of the gondola is covered to avoid the reflection of the sunlight on the inner face of the shields. The electronics boxes are kept on the port side where they are covered from direct sunlight. There, they have a large uncovered portion of the sky to radiate their heat. The primary and secondary mirrors are covered by a cylindrical baffle open on the port side and extending 6~m from the surface of the primary. The design of this baffle required the construction of a stiff, light-weight carbon-fiber and aluminum structure, which is described in Ref. [\citenum{soler2014mecha}]. This solution introduces a small elevation dependence on the azimuth scan ranges, but ultimately allows for Sun avoidance during the observation of the sky between 40\degr\ and 140\degr\ in azimuth, as illustrated in Figure \ref{BLASTpol360}.

\begin{figure}
\centerline{
\vspace{-0.3cm}
\includegraphics[width=0.8\linewidth]{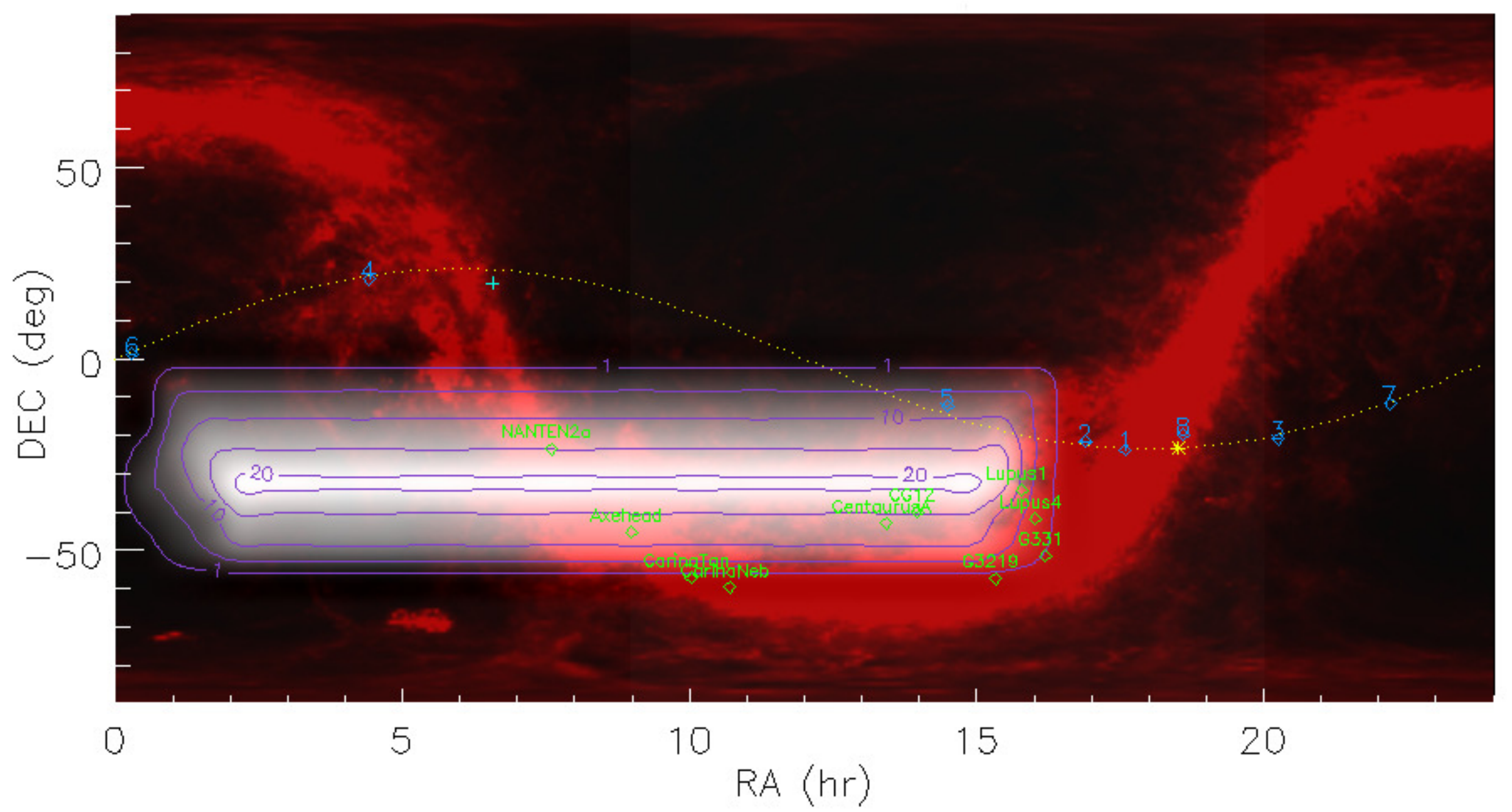}}
\vspace{0.2cm} 
\caption{BLASTPol sky visibility chart corresponding to boresight azimuth range [140\degr,40\degr] with respect to the Sun and [20\degr,50\degr] in elevation for a nominal flight form McMurdo (Lat.~77\degr\ South, Lon.~165\degr\ East) on December 28, 2012. The contours mark observable regions during 1, 5, 10, and 20 hours per day. The background image is the $100\,\mu$m emission measured by {\it IRAS\/} over the whole sky\cite{schlegel1998}. The yellow line corresponds to the trajectory of the Sun in the sky.}\label{BLASTPolObservedSky}
\end{figure}

Figure \ref{BLASTPolObservedSky} shows the observational ranges allowed by the BLASTPol sunshields projected on the sky. On the left of the observable region (shown by the purple contours) are the Lupus I and Lupus IV regions. The thermal model makes possible the assessment of direct sunlight illumination of the primary and secondary mirrors. Additionally it enables evaluation of the temperature changes, which may result from pointing so close to the Sun and which can greatly damage the optical system of the telescope.

In the thermal model the sunshields are simulated by two parallel layers of mylar, separated by a distance equal to the diameter of the tubes in the sunshield frame. Each covered triangle in the sunshield frame is represented by two independent mylar sheets. The sheets are not conductively coupled to each other or to the sunshield frame. Facing outward, the outer layer has the optical properties of Duralar Al Mylar\reg\ (2.0~mil, VDA) and facing the inside of the gondola, the inner layer has the optical properties of Lamart Al Mylar\reg\ (3.0~mil, bonded) as described in Table \ref{OpticalProperties}. The faces of the sunshield that are facing each other are optically active and are simulated as bare aluminum.

\subsection{Telescope Elements}

The BLASTPol telescope elements included in the thermal model are the primary mirror, the telescope struts, the secondary mirror, and the secondary mirror push plate. The primary mirror is simulated by a 1.8-m diameter concave shell, which is $1.0\,$inch thick and is coupled to the inner frame by three aluminum conductors with $(A/x) = 5.0\,$inch, and radiating only on the side facing the sky. This coupling simulates the thermal insulation from the frame by G10 spacers, while the inactive back face corresponds to the 12 layer Multi-Layer Insulation (MLI) added in the gap between the mirror and the frame. The inner frame is simulated by a set of box aluminum beams with one isothermal node each. The nodes are coupled conductively by aluminum and $(A/x)$ value, depending on the length of each segment. Mounted to the inner frame are the gyro box, the star cameras, the Receiver Electronics Crate (REC) and the Data Acquisition System (DAS) crate\cite{galitzki2014}.

The primary mirror is also thermally coupled to the four Carbon Fiber Reinforced Polymer (CFRP) struts covered with thin aluminum fairings. A $(A/x)=1.0\,$inch aluminum coupling between the edge of the primary mirror and the boxes that represent the struts simulates the aluminum inserts, coupling both pieces. The struts are conductively coupled to the push plate by an aluminum conductor with $(A/x)=1.0\,$inch. The push plate has a thermal load of $10.5\,$W, which corresponds to the linear actuators of the automatic focusing system, and is coupled to the secondary mirror by $(A/x)=1.0$~inch aluminum conductor representing the leaf-spring connection.

The modeling of the thermal behavior of large passive elements, such as the primary mirror and the struts, is complicated, because the effect of air coupling becomes relevant. The scope of the thermal model in the case of the optical elements is just to determine the presence of temperature gradients or direct sunlight illumination. These two factors are more critical in the performance of the optical system than the equilibrium temperatures at float. Since the telescope elements are not rated thermally, it is sufficient to let them passively cool down as long as they are thermally stable.

\begin{figure}
\centerline{\includegraphics[angle=270, width=0.45\linewidth]{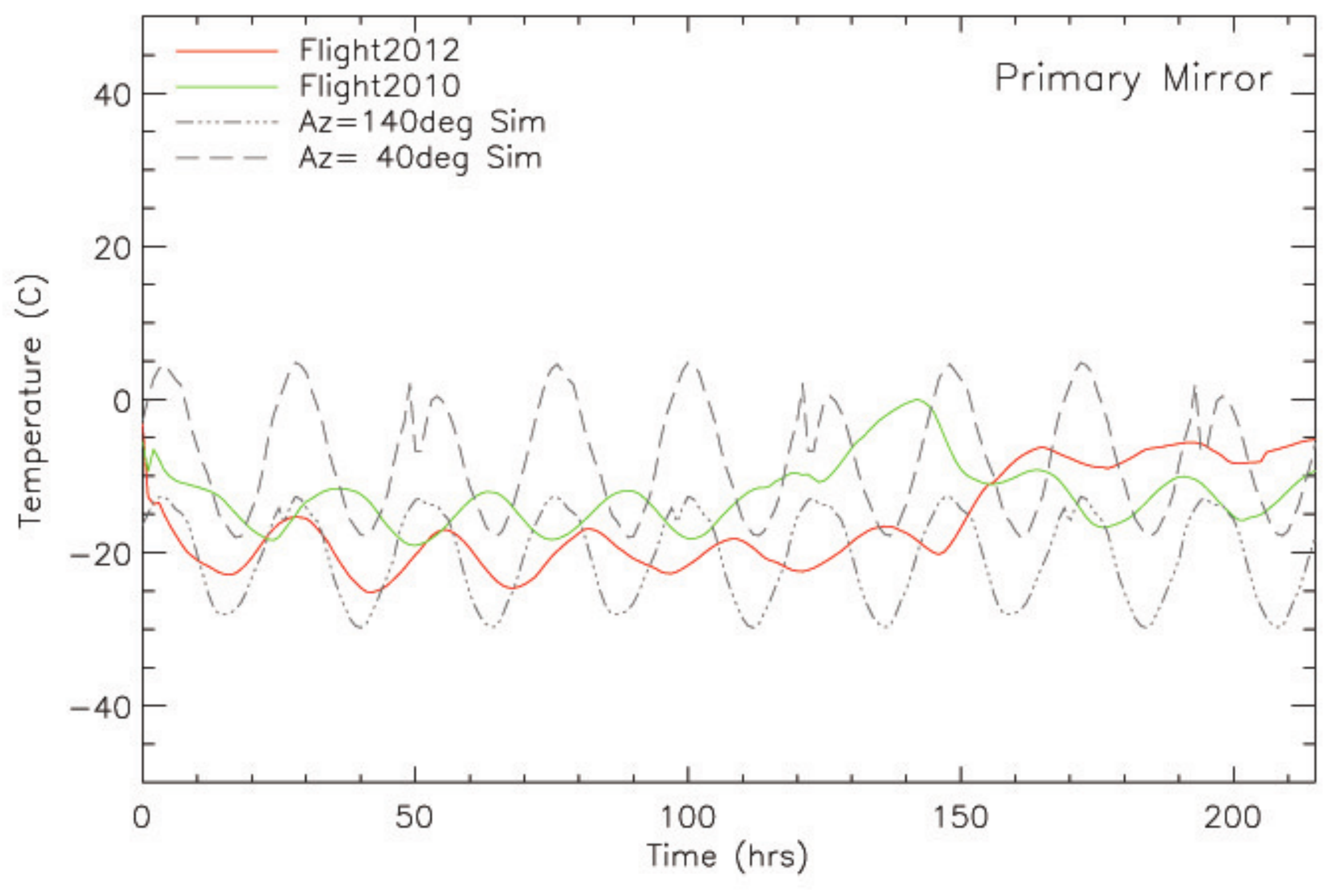}
    \includegraphics[angle=270, width=0.45\linewidth]{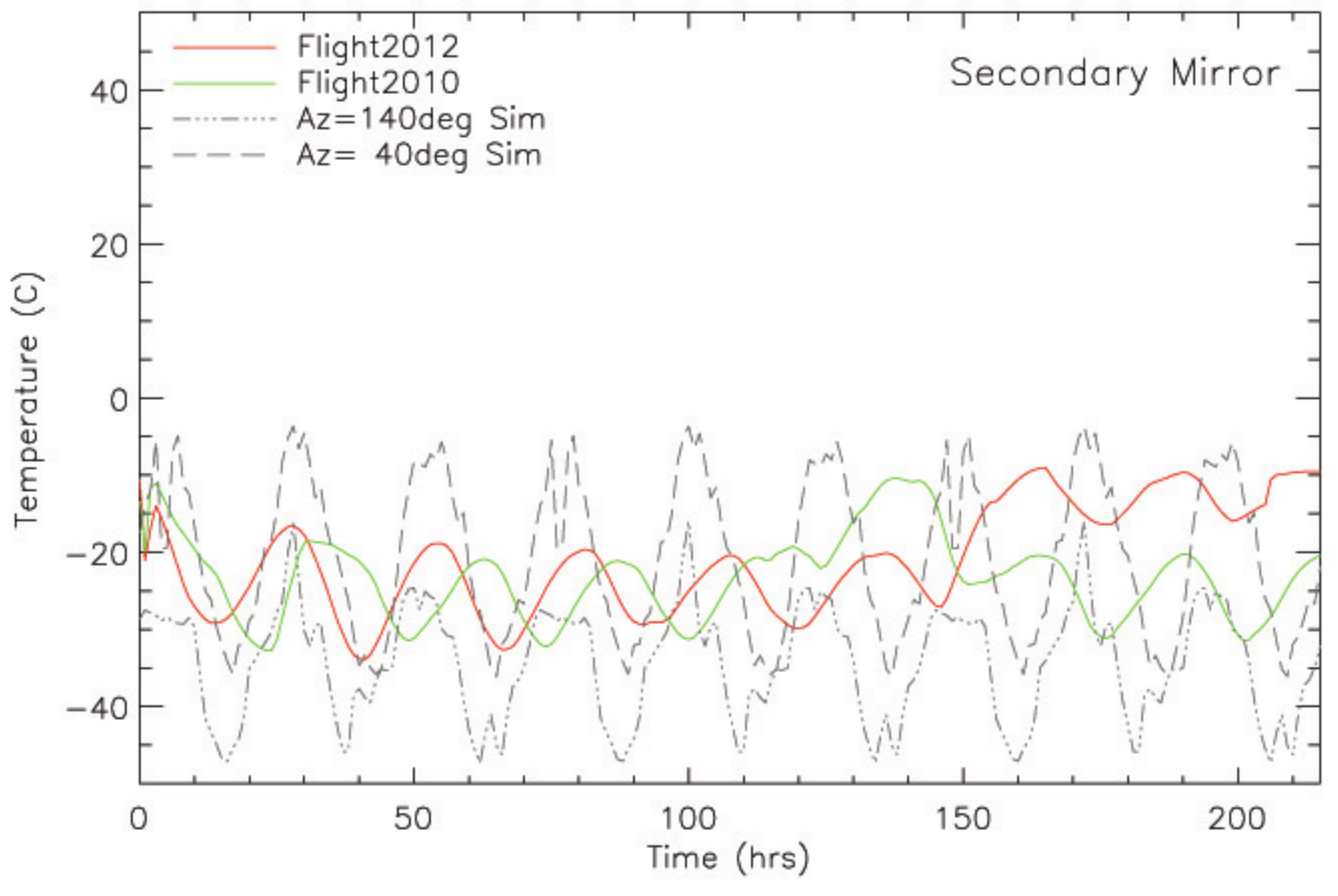}}
\vspace{-0.1cm}
\centerline{  \includegraphics[angle=270, width=0.45\linewidth]{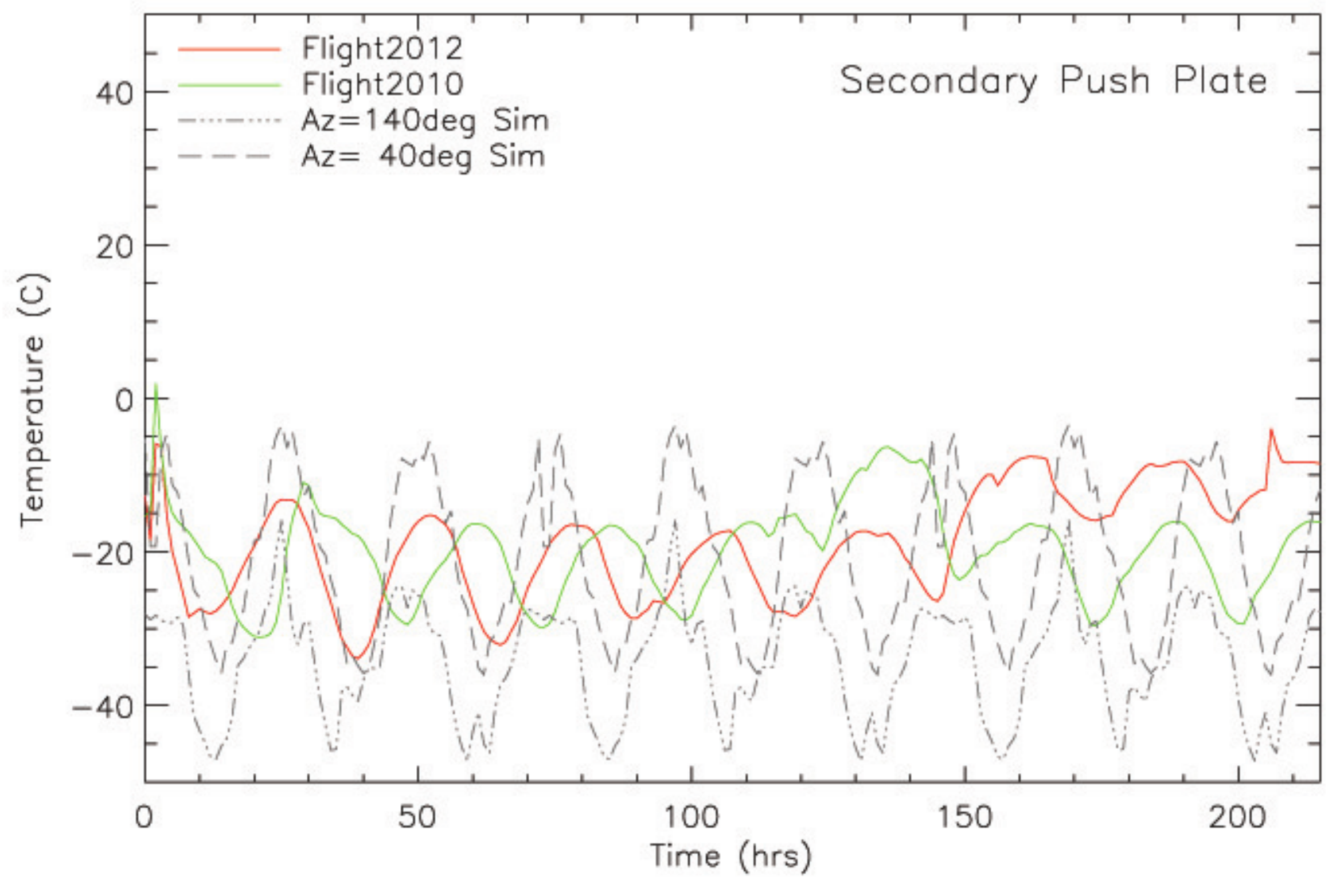}
    \includegraphics[angle=270, width=0.45\linewidth]{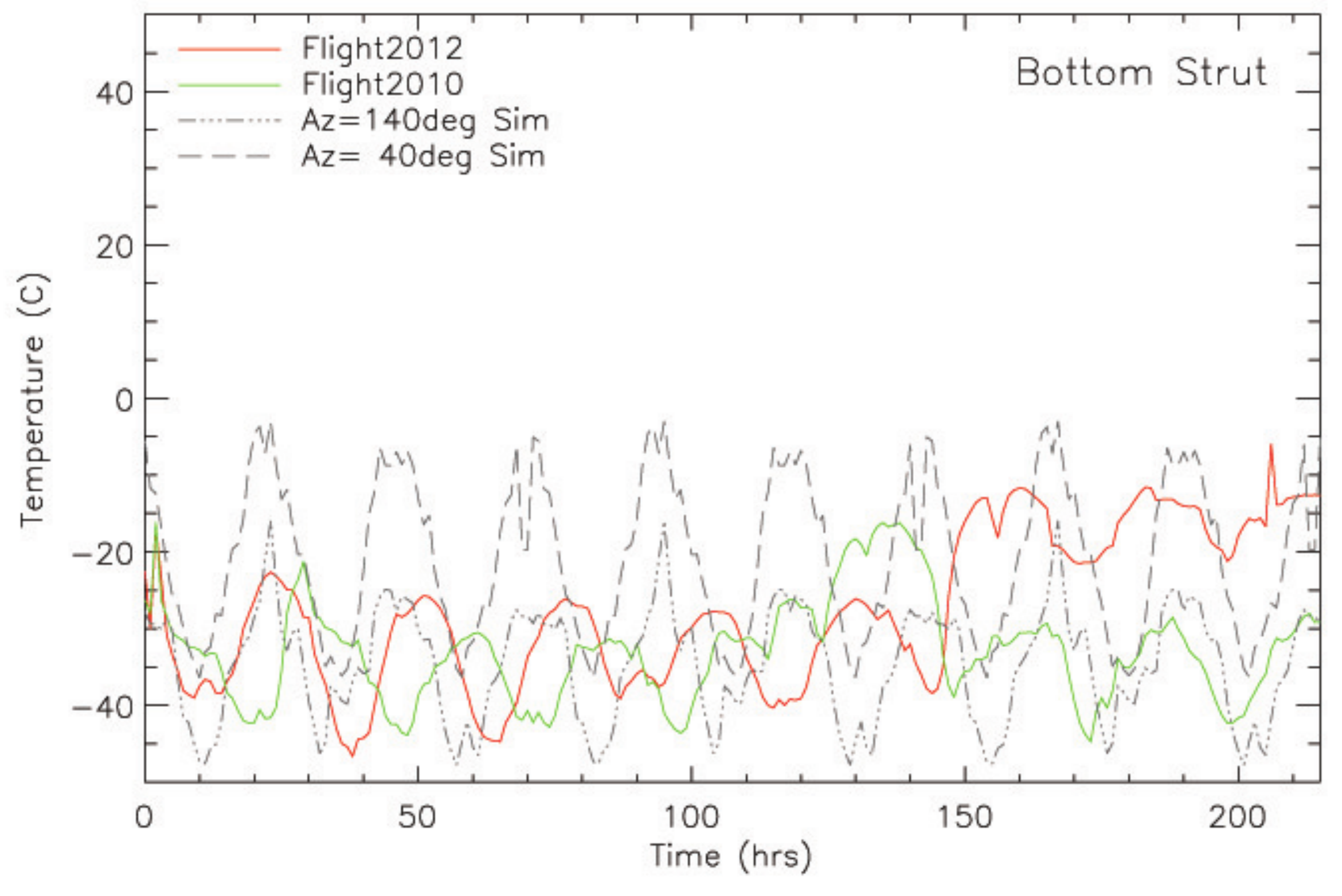}}
    \caption[BLASTPol Outer Frame Electronics Heat Loads and Temperature Ratings]{Flight Temperatures of the BLASTPol telescope elements and predictions of the thermal model. }\label{thermal:blastpoltelescopeTplot}
\end{figure}

Figure \ref{thermal:blastpoltelescopeTplot} shows the in-flight temperatures of the BLASTPol10/12 telescope elements compared to the thermal model predictions and the extreme values summarized in Table  \ref{thermal:blastpolIFelectronicsTtable}. At float, the diurnal temperature changes are $\sim$10\degr\ C for the primary mirror, $\sim$15\degr C for the secondary, and $\sim$20\degr C for the struts. The model is successful in determining the temperature of the primary and the secondary within $\sim$10\degr C. However, the amplitude of the diurnal cycles is larger in the thermal model. This behavior is consistent with air coupling modulating the temperature of the optical elements.

\subsection{Inner and Outer Frame Electronics}

The heat load and temperature ratings of the inner and outer frame electronics are summarized in Table \ref{thermal:blastpolIFelectronicsTtable}. The REC and the ACS crate are mounted to the back beams of the inner frame. Each box is modeled as a $0.125\,$inch wall aluminum box, reproducing the dimensions of the crates they represent. Both sides of the box are optically active, with the inside surface as bare aluminum and the outer face painted white. Each box sits on an inner frame beam and is fastened by aluminum angles, making the conductive coupling approximately $(A/x)=1.0\,$inch aluminum. 

The gyroscopes box (gyrobox) is mounted to one of the beams connecting the front and the back of the inner frame. It is a $0.25\,$inch wall aluminum box with a heat load of $18.36\,$W. It is directly fastened to the beam and this coupling is represented by a $(A/x)=10.5\,$inch aluminum connection. The gyrobox has an internal heating resistor, which is activated when the box temperature is below 0\degr C. The star cameras are mounted in aluminum pedestals on the top front beam of the inner frame. During the BLAST test flight from Fort Sumner, NM in 2003 (BLAST03), the star froze and ceased to work at float, since the outer surface of the cylindrical enclosure had been painted white. The coupling between the star camera enclosures and the inner frame is estimated to be $(A/x)=6.0$~inch aluminum and the heat load on each camera is $28.0\,$W.

\begin{figure}
\centerline{\includegraphics[angle=270, width=0.45\linewidth]{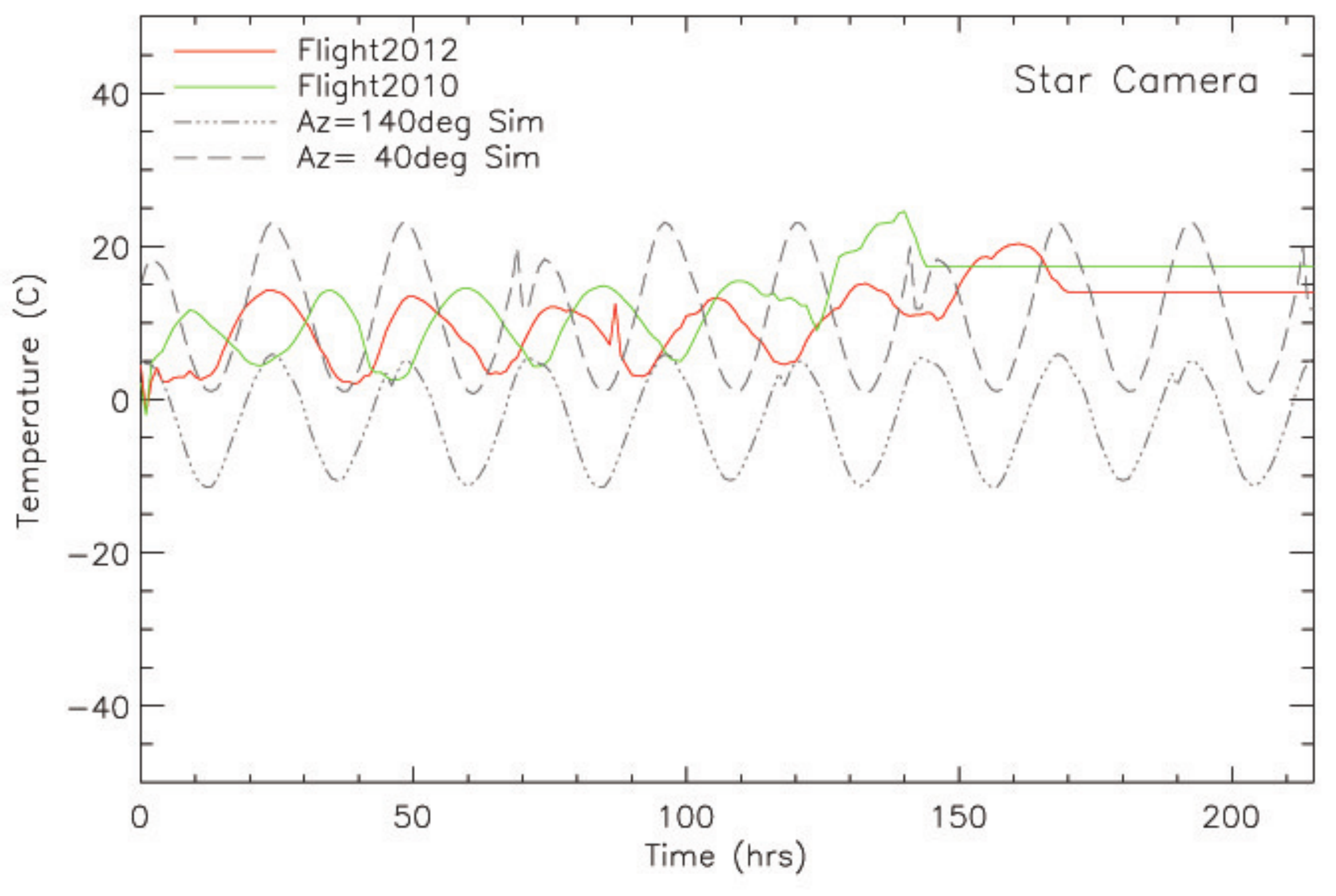}
\includegraphics[angle=270, width=0.45\linewidth]{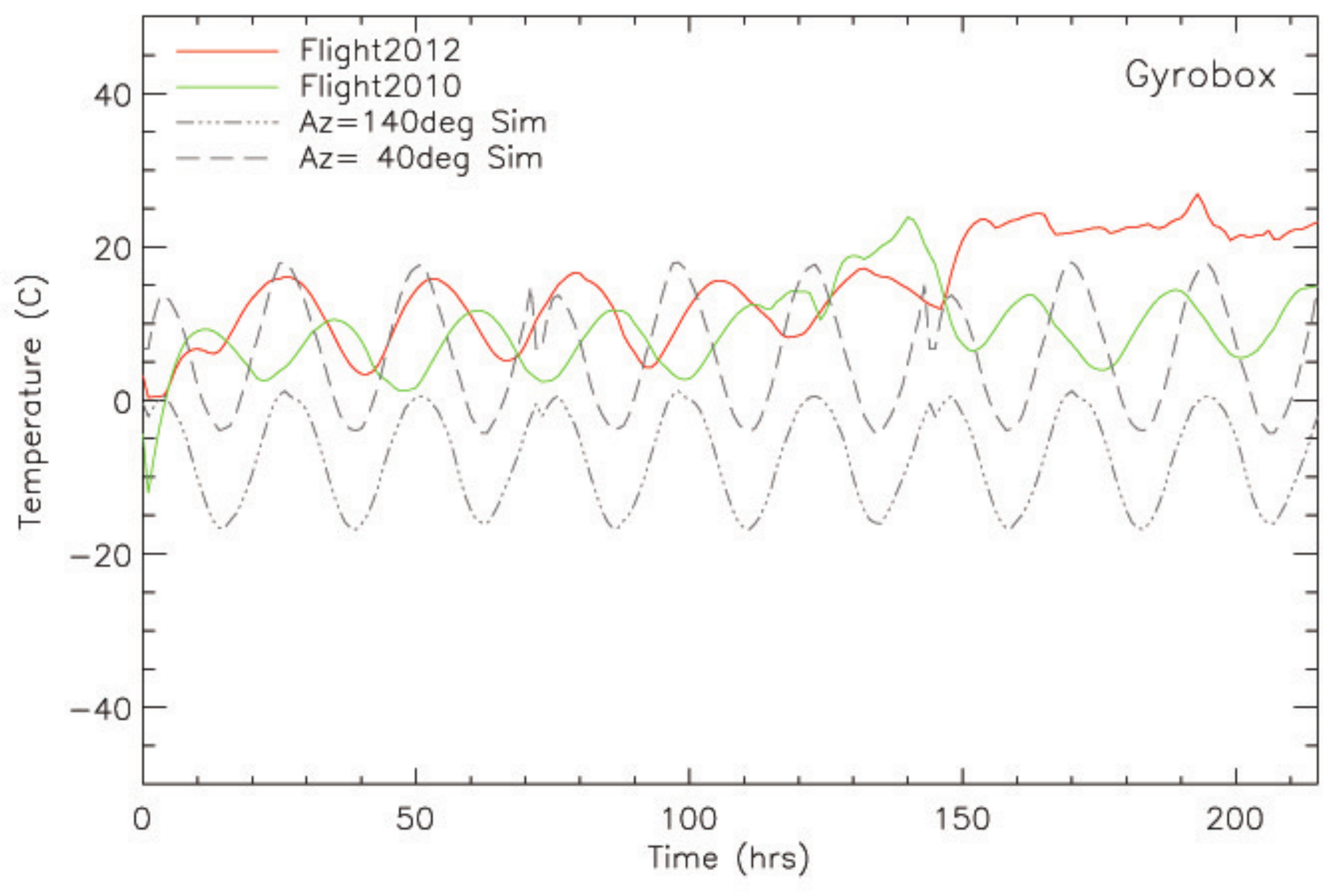}}
\vspace{-0.1cm}
\centerline{\includegraphics[angle=270, width=0.45\linewidth]{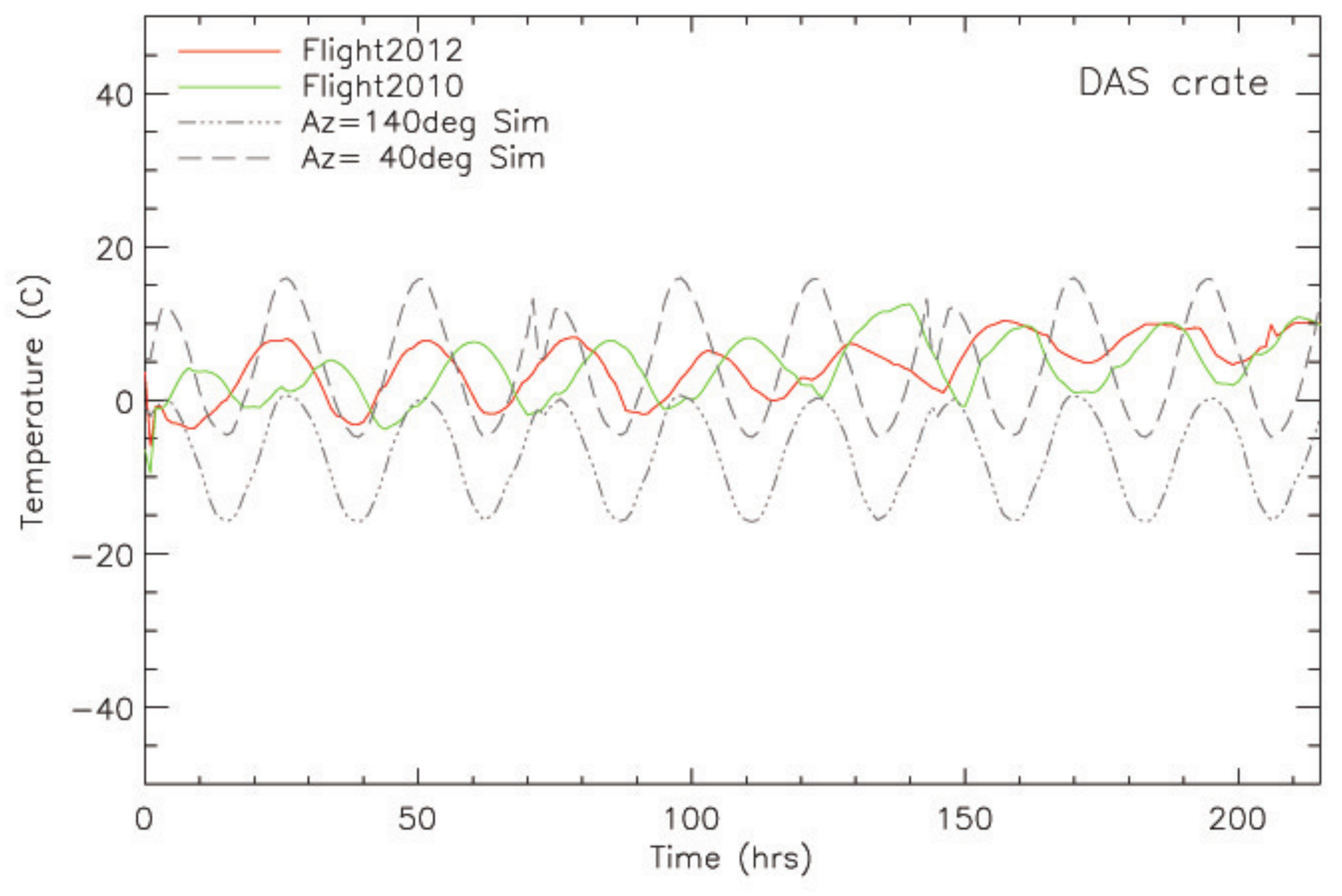}
\includegraphics[angle=270, width=0.45\linewidth]{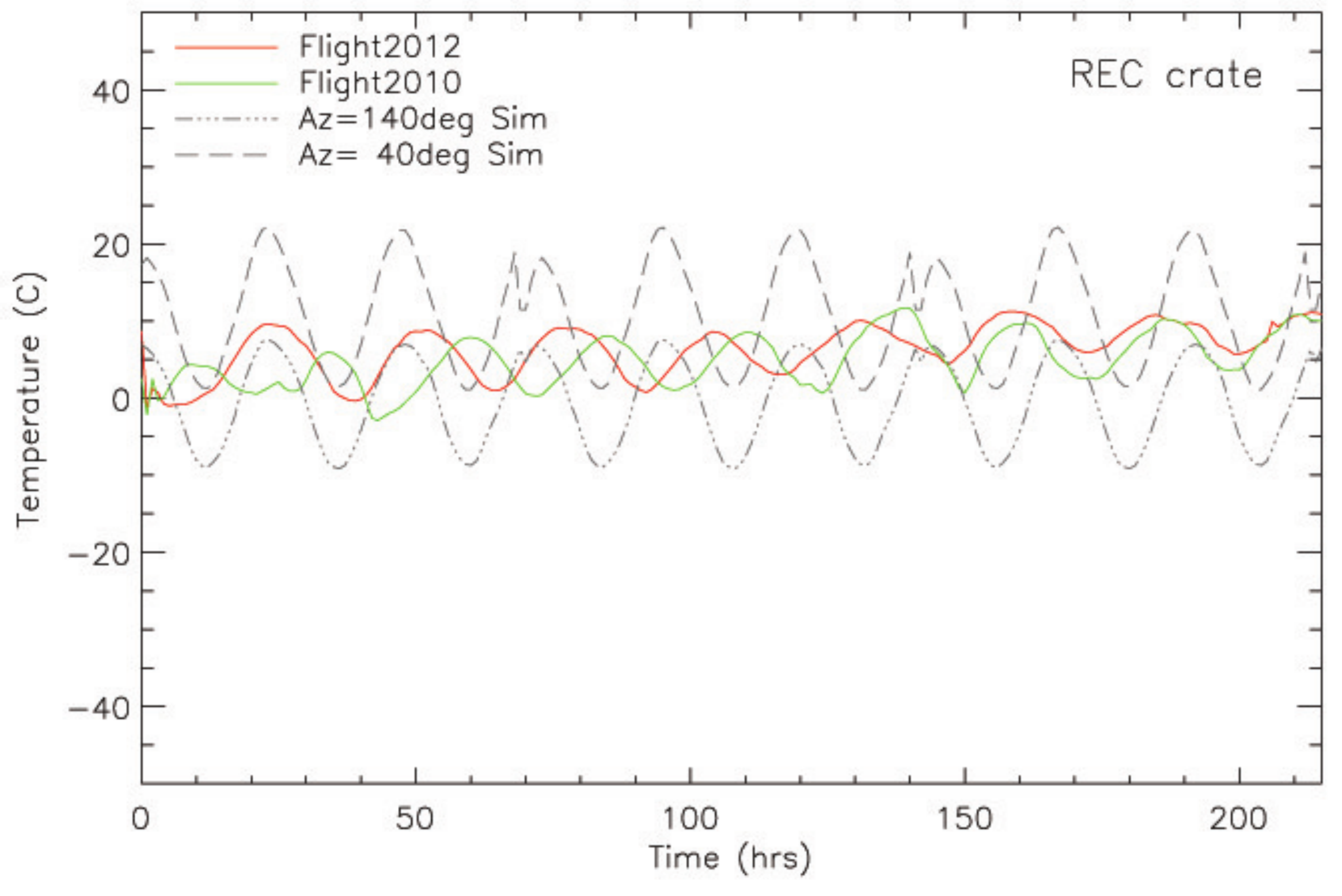}}
\caption[Flight Temperatures of the BLASTPol Inner Frame Electronics]{Flight temperatures of the BLASTPol inner frame electronics.}\label{thermal:blastpolIFelectronicsTplot}
\end{figure}

The temperatures of the inner frame electronics during BLASTPol10/12 are shown in Figure~\ref{thermal:blastpolIFelectronicsTplot} and summarized in Table~\ref{thermal:blastpolIFelectronicsTtable}. All of the electronics boxes operated within the required temperature ranges, and the temperatures predicted by the thermal model are within $\pm$10\degr C of the temperatures at float. There is clearly a more pronounced diurnal temperature change in the simulated temperatures, which is related to the coupling between the electronics boxes and the beams composing the inner frame. However, since the electronics operated far from the temperature limits with the current thermal model configuration, no further correction was considered necessary.

The ACS, the flight computer box, and the serial hub are mounted to a $0.5\,$inch thick aluminum honeycomb platform, which is mounted on the port side of the outer frame. The thermal modeling of the aluminum honeycomb is challenging, since the coupling between both planes is hard to model and depends on the adhesive used to fasten the thin aluminum sheets in the honeycomb with the $6.25\times 10^{-2}\,$inch aluminum slabs and honeycomb pattern itself. The approach followed in the thermal modeling of BLASTPol is to simulate the honeycomb composite as two separate layers of aluminum, coupled radiatively by the inner layers, which have bare aluminum surfaces. Additionally, the conductive coupling is simulated by thermal contact between both surfaces with a material having 1\% the thermal conductivity of aluminum.

The BLASTPol computer box is a descendant of the pressurized enclosure flown in BLAST03\cite{pascale2008}. The advent of Solid-State Disks (SSDs) permits the use of non-pressurized enclosures and the BLASTPol computer box is a compact solution to store the SSD and the two Motherboards. However, this compact model does not have enough surface area to dissipate the heat generated by the chips in the motherboard and for this reason, two copper braid straps are mounted directly to the aluminum heat sinks on top of the motherboards. The other end of the straps is connected to the gondola frame. The heat load on the chips is estimated to be $15\,$W, and the thermal coupling provided by the heat straps is $(A/x) = 6\,$inch. The serial hub is modeled as a $0.125\,$inch aluminum box with a heat load of $5\,$W. It is mounted to the Hexcel\reg\ desk by aluminum angle beams with a $(A/x) = 5\,$inch aluminum coupling.

\begin{figure}
\centerline{\includegraphics[angle=270, width=0.45\linewidth]{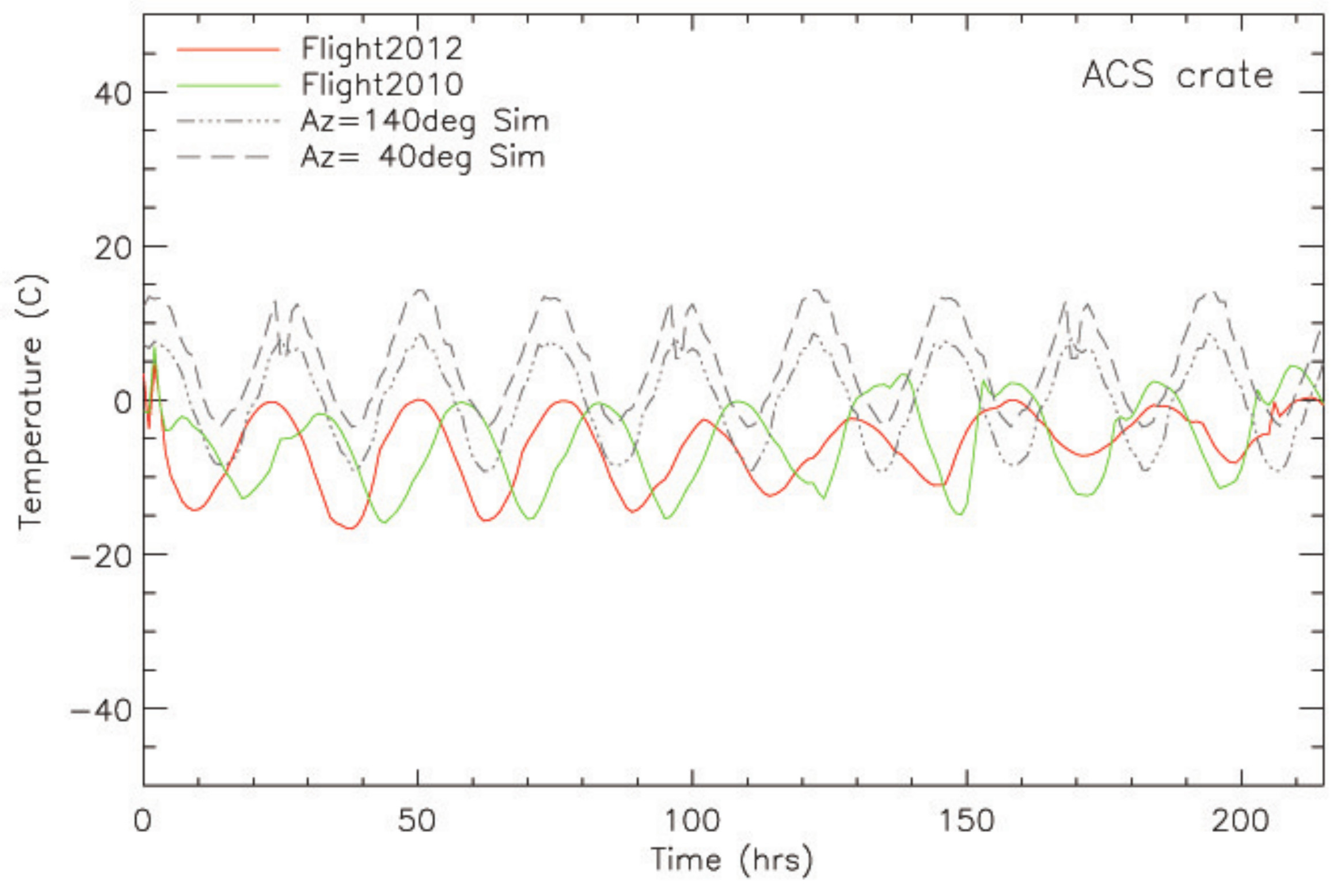}
    \includegraphics[angle=270, width=0.45\linewidth]{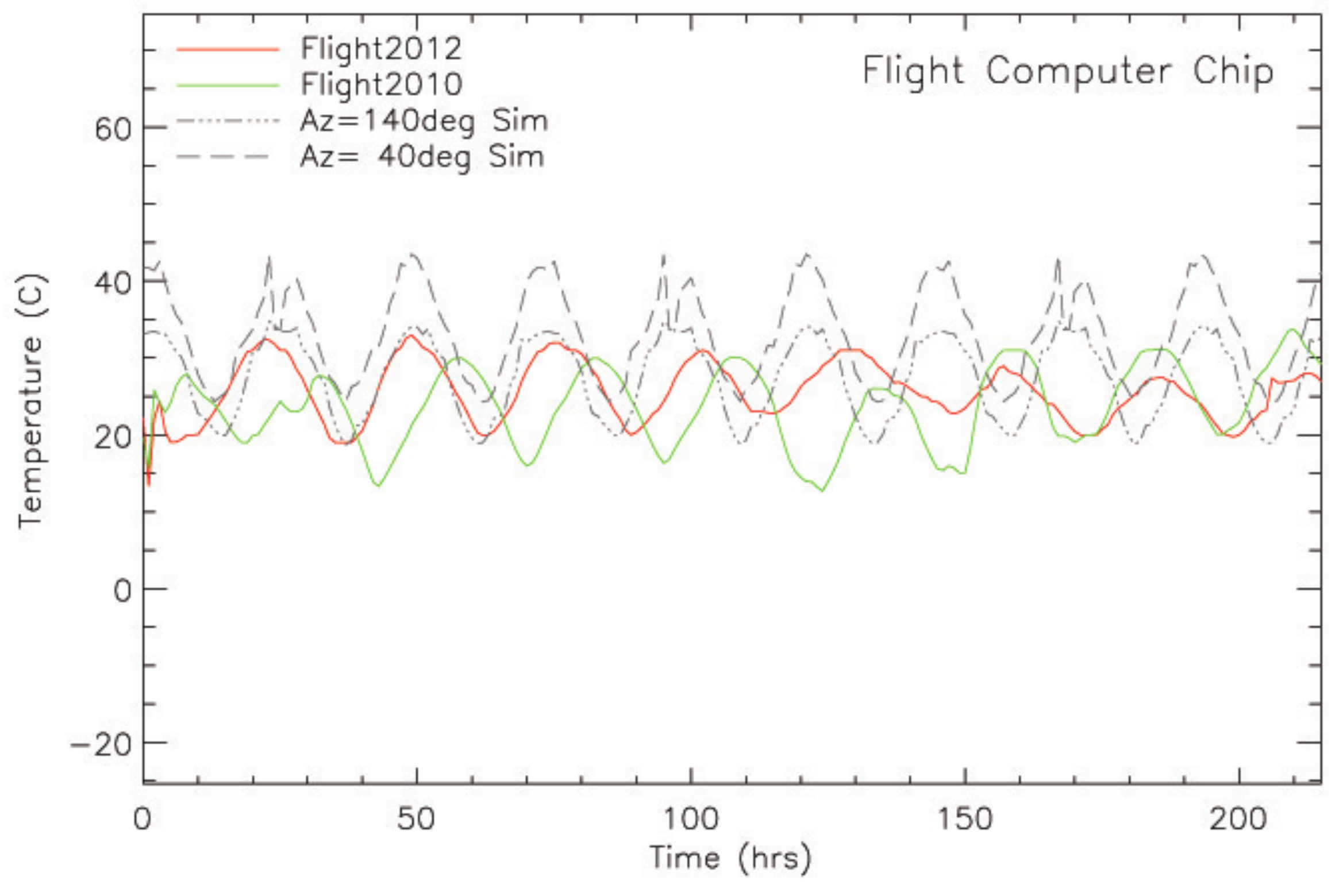}}
\vspace{-0.1cm}
\centerline{\includegraphics[angle=270, width=0.45\linewidth]{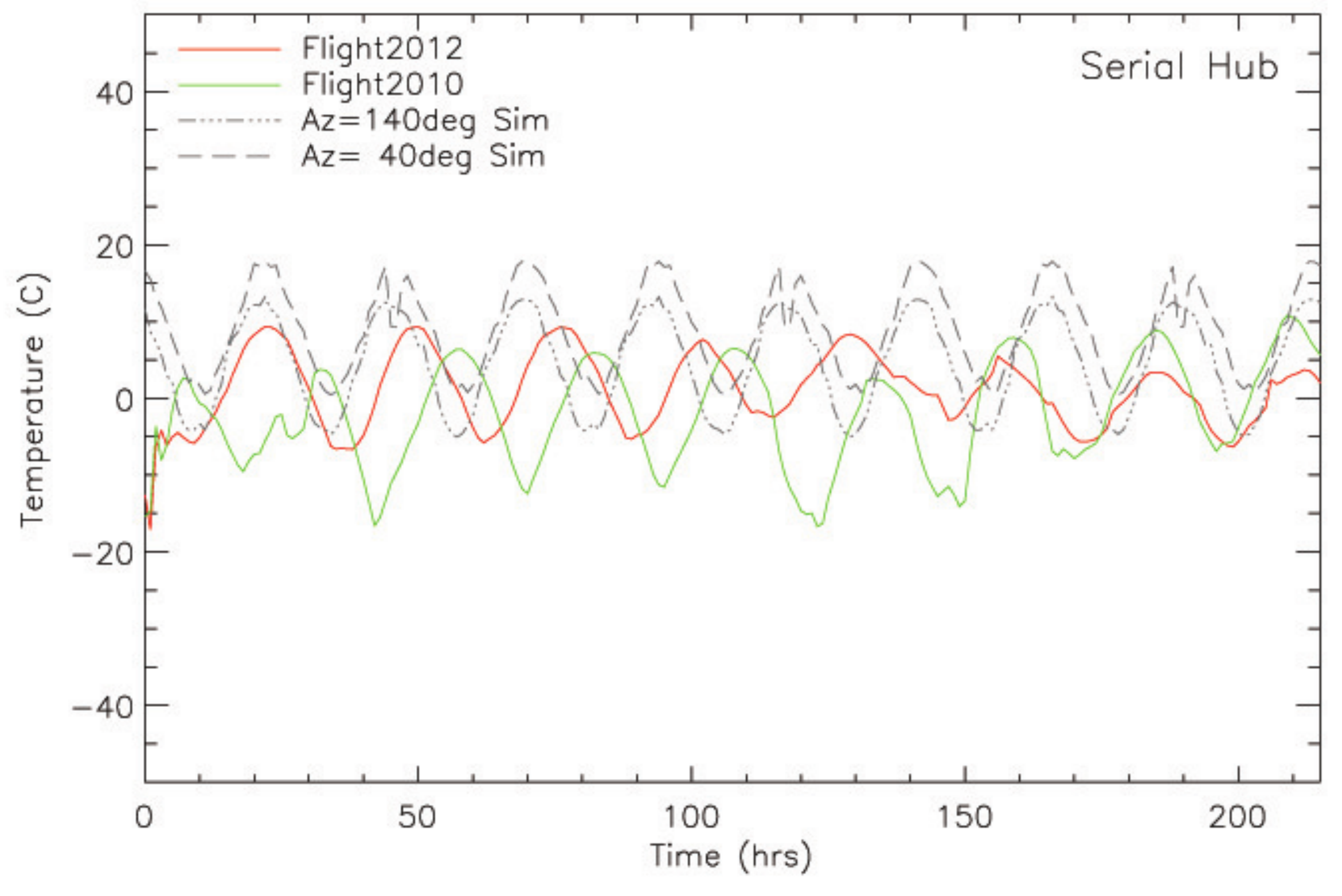}
    \includegraphics[angle=270, width=0.45\linewidth]{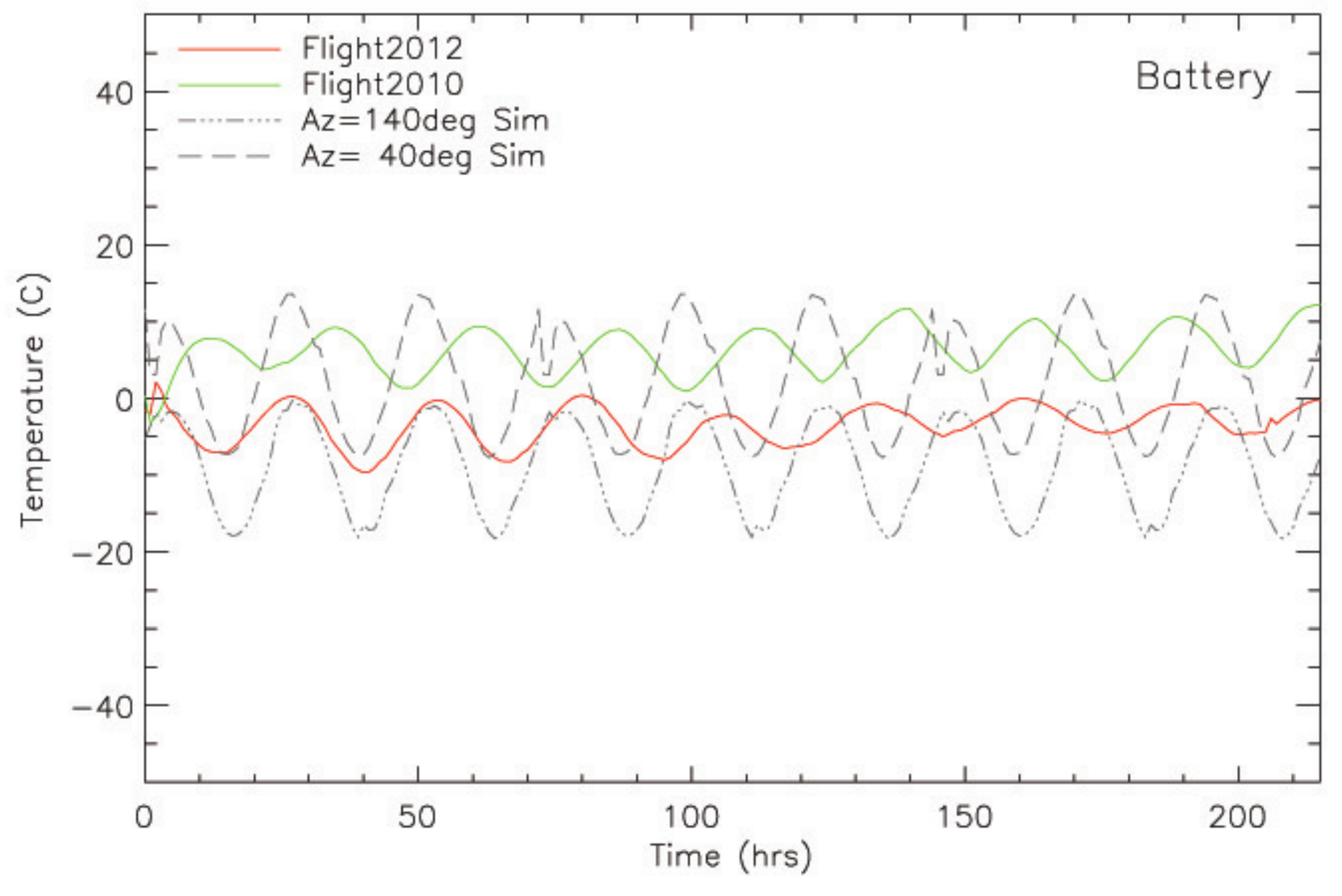}}
  \caption[Flight Temperatures of the BLASTPol Outer Frame Electronics]{Flight temperatures of the BLASTPol outer frame electronics.}\label{thermal:blastpolOFelectronicsTplot}
\end{figure}

The temperatures of the outer frame electronics during the BLASTPol10/12 are shown in Figure \ref{thermal:blastpolOFelectronicsTplot} and summarized in Table \ref{thermal:table1}. All of the electronics boxes operated within the adequate temperature ranges and the temperatures predicted by the thermal model are within $\pm$10\degr C of the temperatures at float. The diurnal temperature change is larger in the simulated data, as discussed in the case of outer frame electronics. The ACS crate and the serial hub reached minimum temperatures 10\degr C lower than predicted by the model, although these are 15\degr C over the minimum temperature of operation. The source of this discrepancy is possibly that the inner surface of both boxes is emissive and these were non-active in the model. In BLASTPol12 an attempt to reduce the emissivity of the ACS was made by partially covering one of its external faces with aluminum tape, but this proved to have no effect on the temperatures at float, as shown in Figure~\ref{thermal:blastpolOFelectronicsTplot}.

\begin{table}[htbp!]
\caption{Heat loads and temperature ratings of BLASTPol electronics and telescope elements. The maximum and minimum temperatures of operation were determined during vacuum and cold air tests made at CSBF in the summer of 2010. The minimum in-flight temperatures correspond to the values measured at float.}\label{thermal:blastpolIFelectronicsTtable}
\begin{center} 
  \begin{tabular}{|l|c|c|c|c|c|c|c|}
    \hline
   Component    & Heat load & \multicolumn{6}{|c|}{Temperature Range (\degr C)}\\
                & (W)       & \multicolumn{2}{|c|}{Rating} & \multicolumn{2}{|c|}{Model} & \multicolumn{2}{|c|}{Flight} \\
                &           & Min   & Max   & Min   & Max  & Min   & Max \\ \hline
   Primary mirror           & 0         & $\dots$   & $\dots$   & $-$29.8 & 4.8   & $-$25.2 & 1.4 \\
   Secondary mirror         & 0         & $\dots$   & $\dots$   & $-$47.3 & $-$3.6  & $-$34.1 & $-$9.0 \\
   Push plate               & 10.5         & $\dots$   & $\dots$   & $-$47.2 & $-$3.5  & $-$33.9 & 1.9 \\
   Strut bottom             & 0         & $\dots$   & $\dots$   & $-$47.8 & $-$3.0  & $-$46.6 & $-$5.0 \\   \hline 
  DAS    & 35        & $-$20   & 40   & $-$15.8 & 15.9  & $-$9.5 & 12.5 \\
   REC    & 65        & $-$20   & 40   & $-$9.1  & 22.2  & $-$2.9 & 11.7 \\
   Gyrobox      & 18.36     & $-$20   & 40   & $-$16.8 & 17.9  & $-$12.1 & 26.8 \\
   ISC          & 28        & $-$27   & 40   & $-$11.4 & 23.1  & $-$2.1 & 24.5 \\
   OSC          & 28        & $-$27   & 40   & $-$11.4 & 23.1  & $-$2.0 & 24.9 \\ \hline
   ACS                & 27        & $-$33    & 55   & $-$9.2 & 14.3   & $-$16.7 & 6.9 \\
   Battery (4)              & 50        & $-$40   & 40   & $-$18.3 & 13.6  & $-$9.7 & 12.2 \\
   Flight computers box     & 20        & $-$31   & 54   & $-$16.8 & 17.9  & $-$12.1 & 26.8 \\
   CPU(2)                   & 15        & 2     & 70   & 18.7 & 43.5  & 12.7 & 33.7 \\
   Serial hub               & 5         & $-$31   & 54   & $-$5.0 & 17.8  & $-$17.1 & 10.8 \\ \hline
 \end{tabular}
\end{center}
\end{table}

\subsection{Motion Systems}\label{thermal:blastpolmotionsystems}

The pivot, the elevation drive, and the reaction wheel motors are the elements of the BLASTPol gondola with higher power consumption, therefore their thermal performance is critical for the operation of the experiment. The estimated heat loads and thermal ratings of these elements are summarized in Table~\ref{thermal:table1}.

\begin{figure}
\centerline{\includegraphics[angle=270, width=0.45\linewidth]{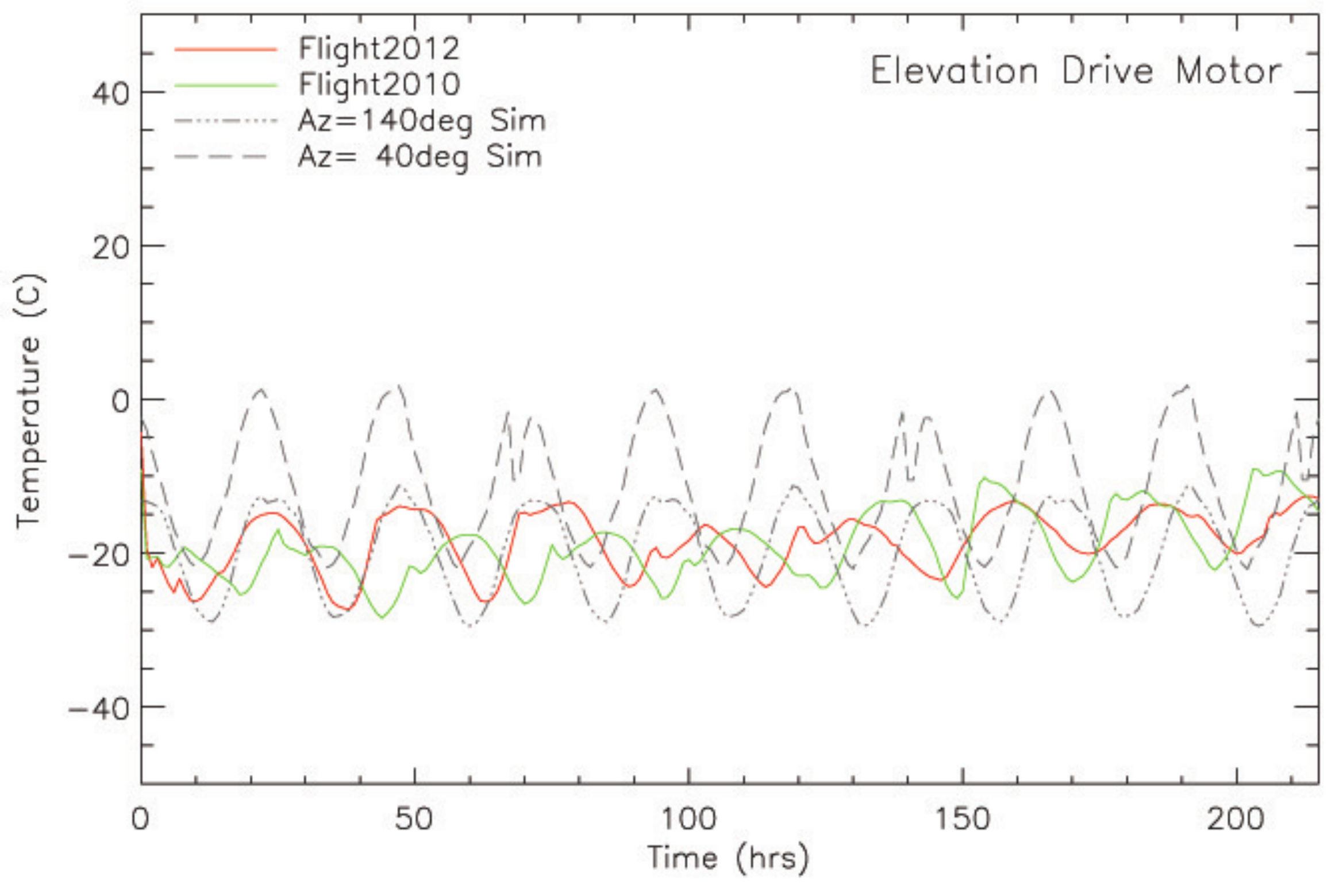}
\includegraphics[angle=270, width=0.45\linewidth]{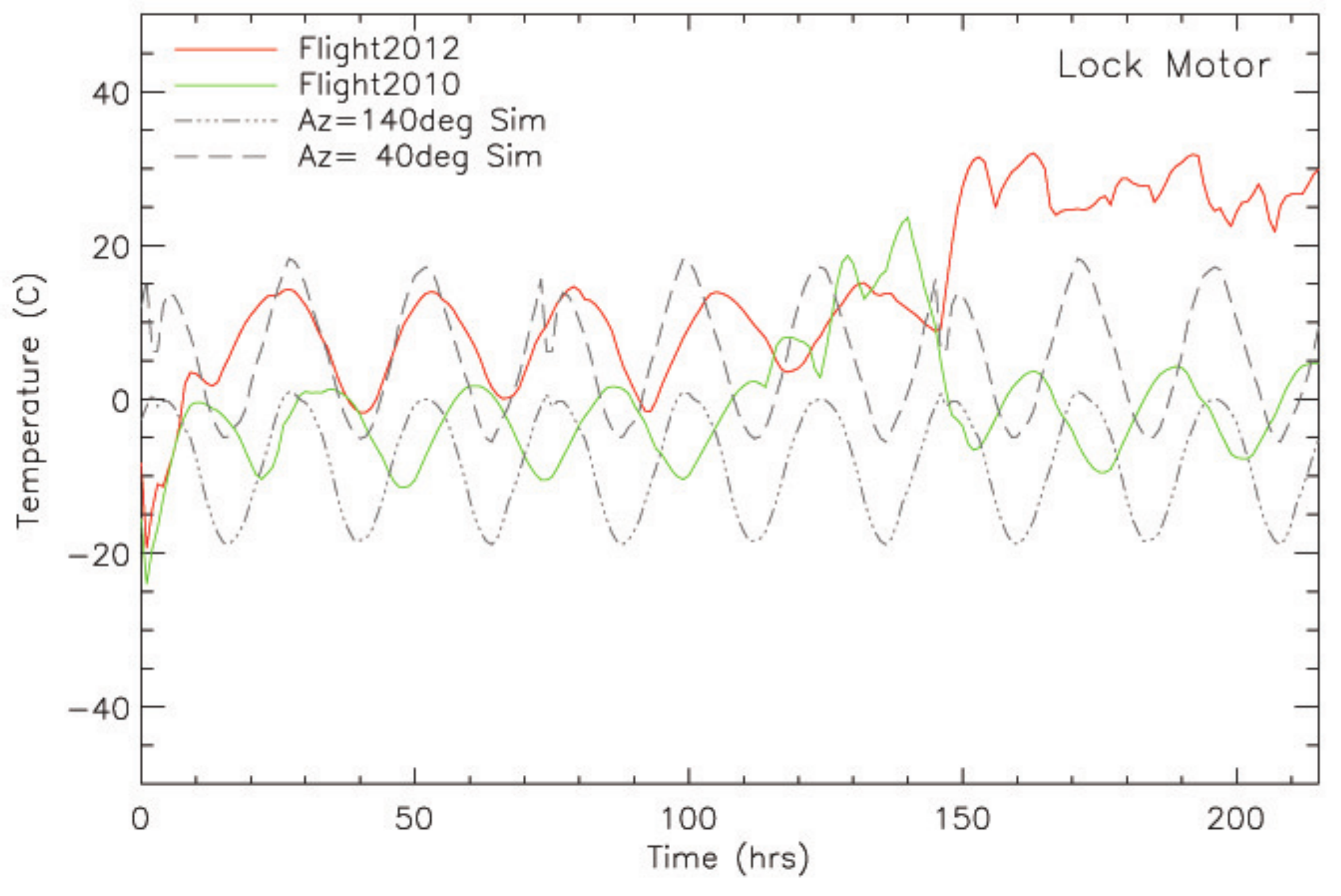}}
\vspace{-0.1cm}
\centerline{\includegraphics[angle=270, width=0.45\linewidth]{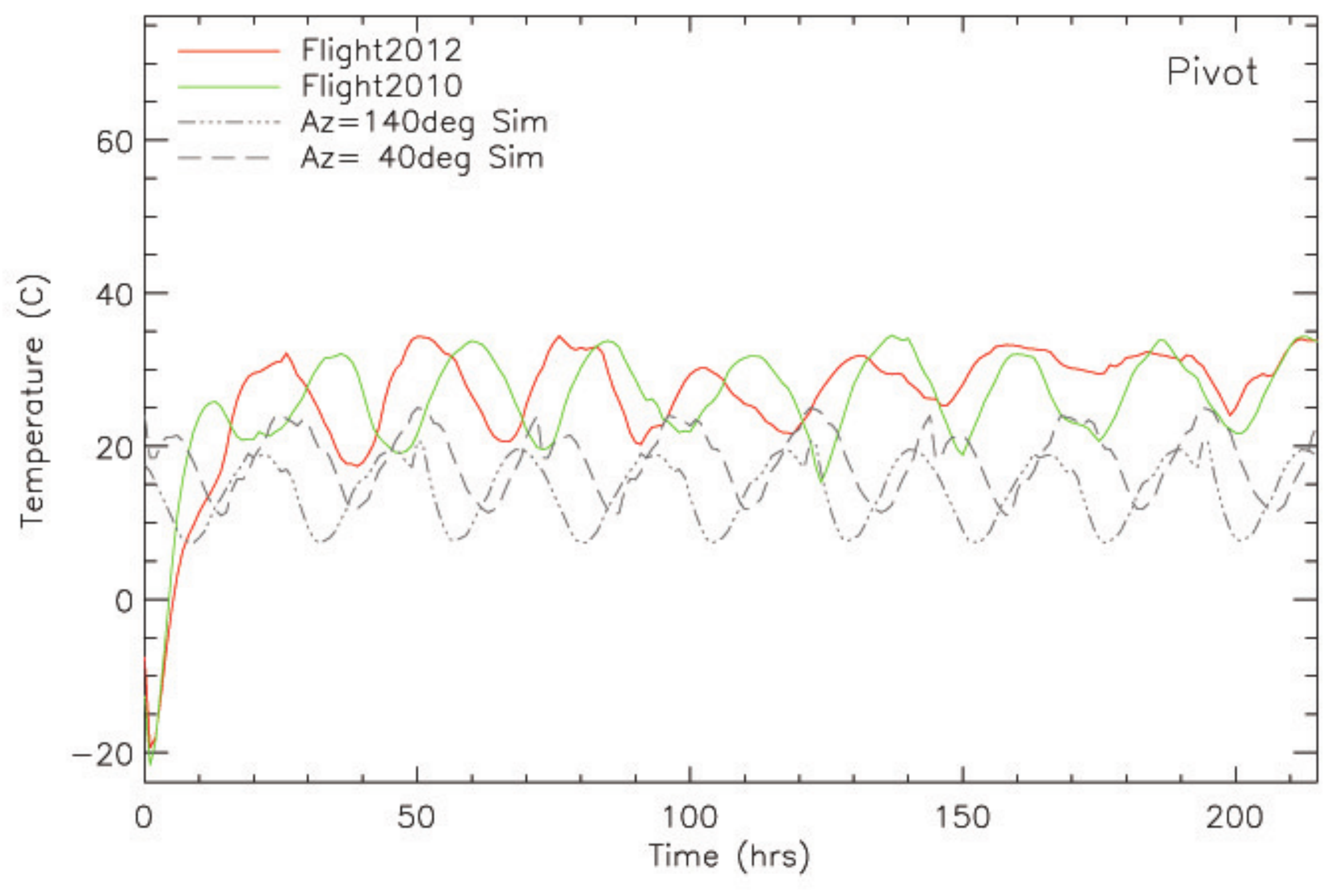}
\includegraphics[angle=270, width=0.45\linewidth]{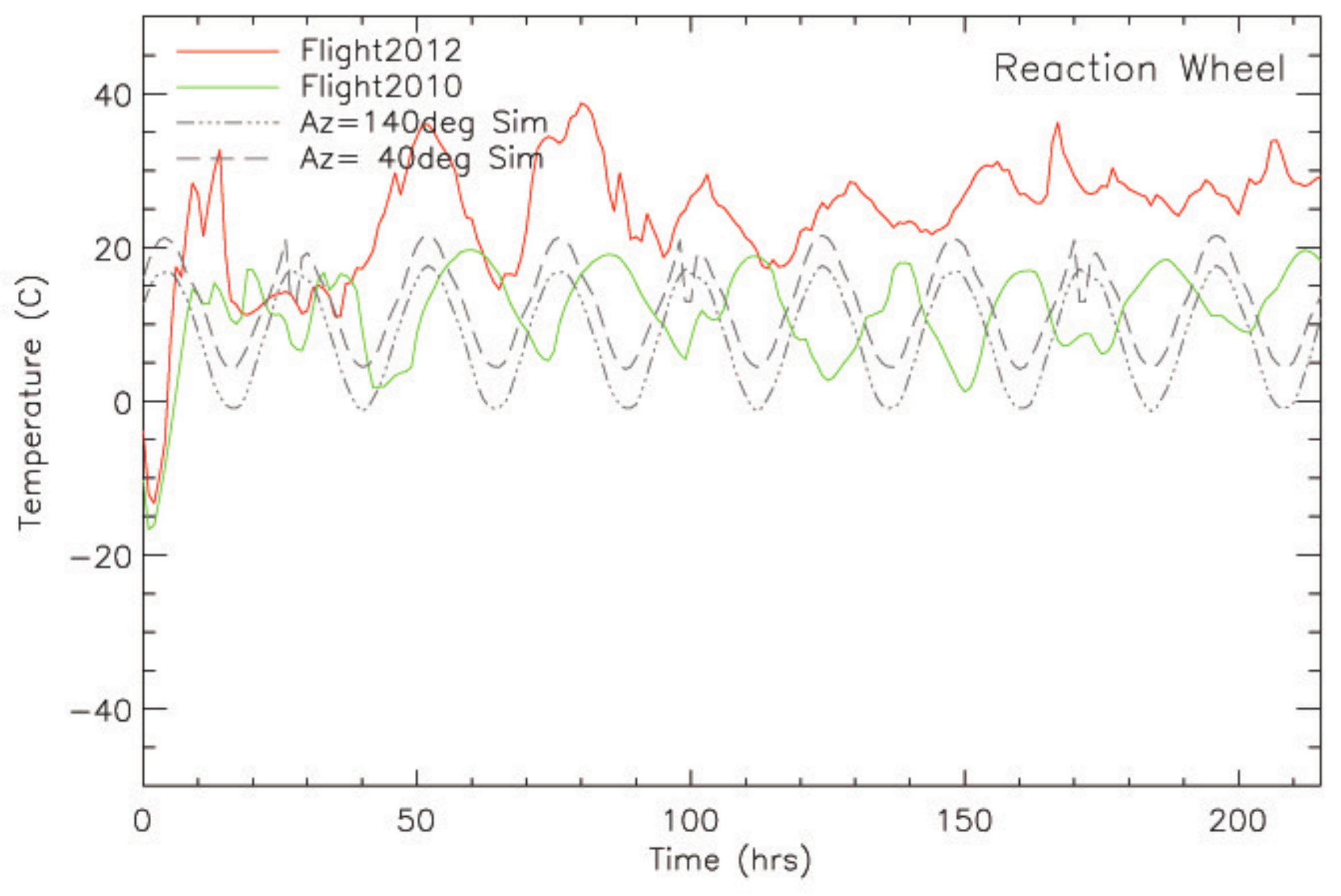}}
\caption[Flight Temperatures of the BLASTPol Motion Systems]{Flight temperatures of the BLASTPol motion systems.}\label{thermal:blastpolmotorsTplot}
\end{figure}

The elevation drive is simulated as a steel cube with a heat load of $6.73\,$W, corresponding to the root mean-square (RMS) current measured during gondola pointing tests in the laboratory. The lower temperature of the motor, $\sim$30\degr C was correctly estimated, while the higher temperature limit was correct for the azimuth position at 140\degr\ from the Sun. The reaction wheel motor is simulated as a steel cylinder with a heat load of $42\,$W, corresponding to the RMS current during gondola pointing tests in the laboratory. During BLASTPol10, the reaction wheel suffered damage and after the second day of flight it was unusable. The temperatures registered during this flight, shown in Figure \ref{thermal:blastpolmotorsTplot}, indicate that the damage was not caused by a thermal malfunction. During BLASTPol12, the reaction wheel operated normally, and the RMS current corresponds to a heat load of $46.1\,$W.

The pivot is the element of the gondola that presents more complications in the thermal modeling, as shown also in Figure~\ref{thermal:blastpolmotorsTplot}. Located on top of the sunshields, the pivot is directly exposed to the sunlight. It is simulated as a steel cylinder with a heat load of $10.4\,$W corresponding to the 80\% of RMS current measured during gondola pointing tests in the laboratory. Beneath the pivot is the controller box, which is modeled as a $1/8\,$inch thick aluminum box covered with silver Teflon tape. The heat load on this box is $2.6\,$W, resulting from the 20\% of the RMS corresponding current.

During BLASTPol10 the pivot temperature peaked at 34.5\degr C, approximately 10\degr C hotter than predicted by the model. This discrepancy can have three main sources: (1) increased power consumption due to its use as the main pointing mechanism after the failure of the reaction wheel; (2) kept in the Sun and with a small surface area, the model of the pivot could underestimate its effective radiative surface area; and (3) the coupling to the suspension cables and the universal joint could increase the temperature of the pivot because these elements have a high $\alpha/\epsilon$ and can become warm when directly exposed to the sunlight.

\begin{table}[htbp!]
\caption{Heat loads and temperature ratings of BLASTPol inner frame electronics. The maximum and minimum temperatures of operation were determined during vacuum and cold air test made at CSBF in the summer of 2010.}\label{thermal:table1}
\begin{center} 
  \begin{tabular}{|l|c|c|c|c|c|c|c|}
    \hline
   Component    & Heat load & \multicolumn{6}{|c|}{Temperature Range (\degr C)}\\
                & (W)       & \multicolumn{2}{|c|}{Rating} & \multicolumn{2}{|c|}{Model} & \multicolumn{2}{|c|}{Flight} \\
                &           & Min   & Max   & Min   & Max  & Min   & Max \\ \hline
   Elevation Drive                  & 5.6       & $-$3.0  & 29.0  & $-$29.5 & 1.8   & $-$28.4 & $-$4.3 \\
   Reaction Wheel                   & 14.24     & $-$4.0  & 32.0  & $-$1.3 & 21.5  & $-$16.7 & 38.7 \\
   Pivot                            & 10.4      & $-$33.0 & 48.0  & 7.4 & 25.0 & $-$21.6 & 34.5 \\
   Pivot Motor Controller           & 2.6       & $-$22.0 & 44.0  & 7.4 & 24.9 & $-$24.4 & 35.1 \\   
   Lock Motor                       &           & $-$27.0 & 51.0  & $-$18.9 & 18.3  & $-$24.1 & 31.9 \\  \hline
   Chin                             & $\dots$     & $\dots$     & $\dots$     & $-$46.5 & 27.7  & $-$48.1 & 10.4 \\
   Gondola frame port back          & $\dots$     & $\dots$     & $\dots$     & $-$5.4  & 13.7  & $-$25.3 & 29.1 \\
   Gondola frame starboard front    & $\dots$     & $\dots$     & $\dots$     & $-$0.1  & 24.5  & $-$21.9 & 28.0 \\
   Hexcel deck top                  & $\dots$     & $\dots$     & $\dots$     & $-$2.7  & 20.4  & $-$16.9 & 12.9 \\
   Hexcel deck bottom               & $\dots$     & $\dots$     & $\dots$     & $-$2.6  & 20.5  & $-$21.7 & 17.8 \\
   Solar Array                      & $\dots$     & $\dots$     & 100   & 39.8  & 107.6 & $-$24.7 & 87.1 \\
   Rear sunshield                   & $\dots$     & $\dots$     & $\dots$     & $-$21.2 & 5.3   & $-$32.4 & 47.7 \\
   \hline
  \end{tabular}
\end{center}
\end{table}

\subsection{Passive Elements and Solar Array}

\begin{figure}
\begin{center}
\centerline{
    \includegraphics[angle=270, width=0.45\linewidth]{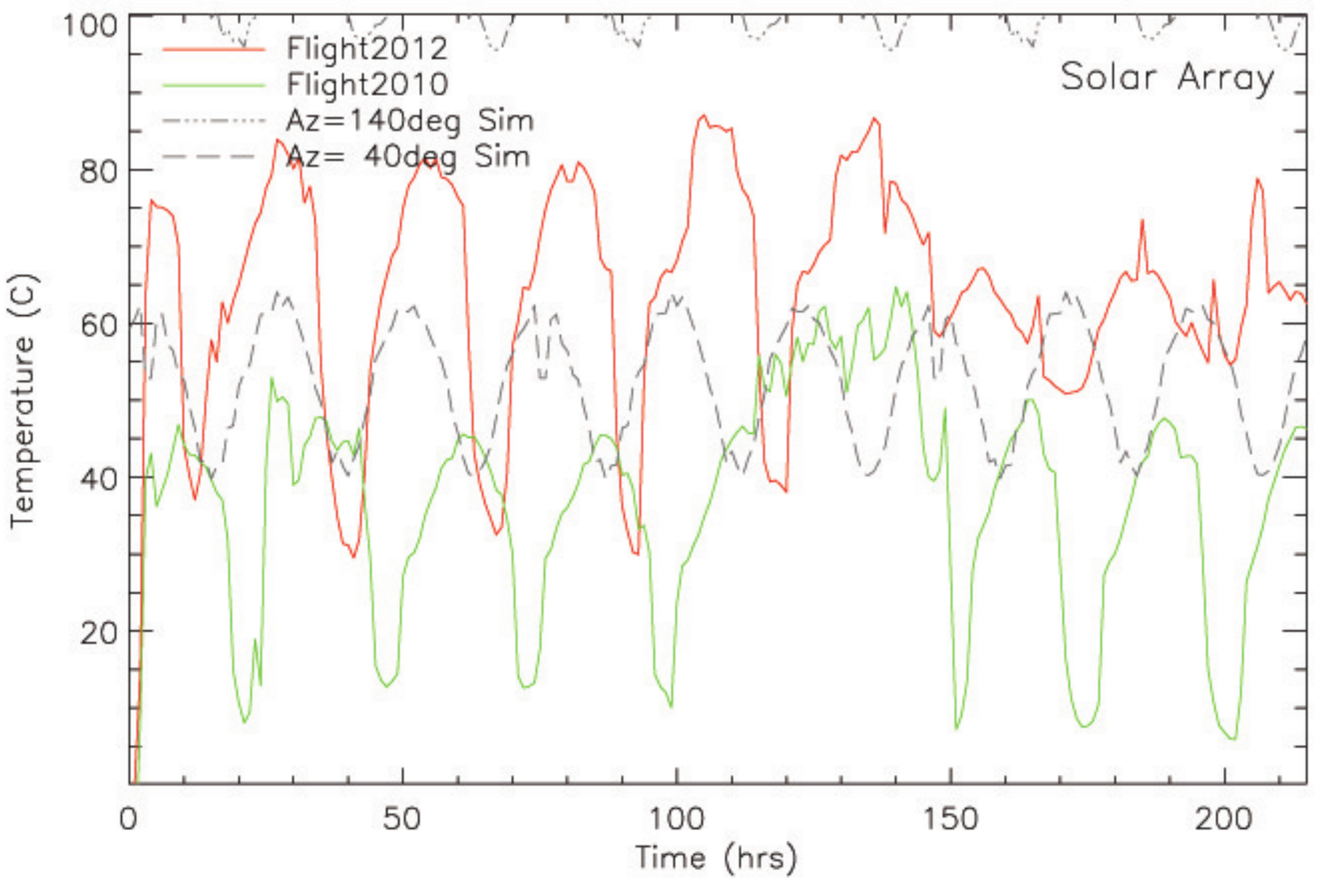}
    \includegraphics[angle=270, width=0.45\linewidth]{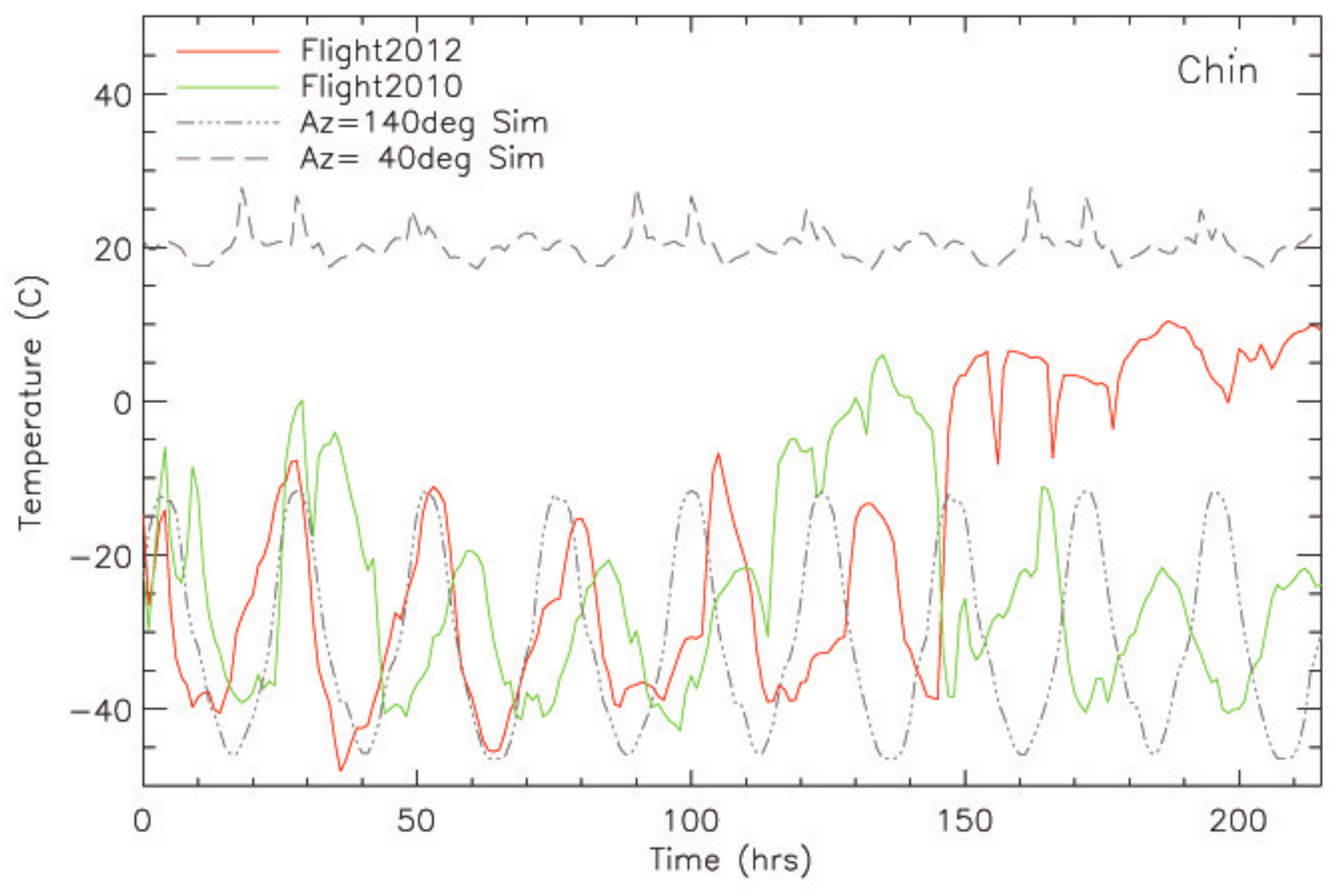}}
\centerline{
    \includegraphics[angle=270, width=0.45\linewidth]{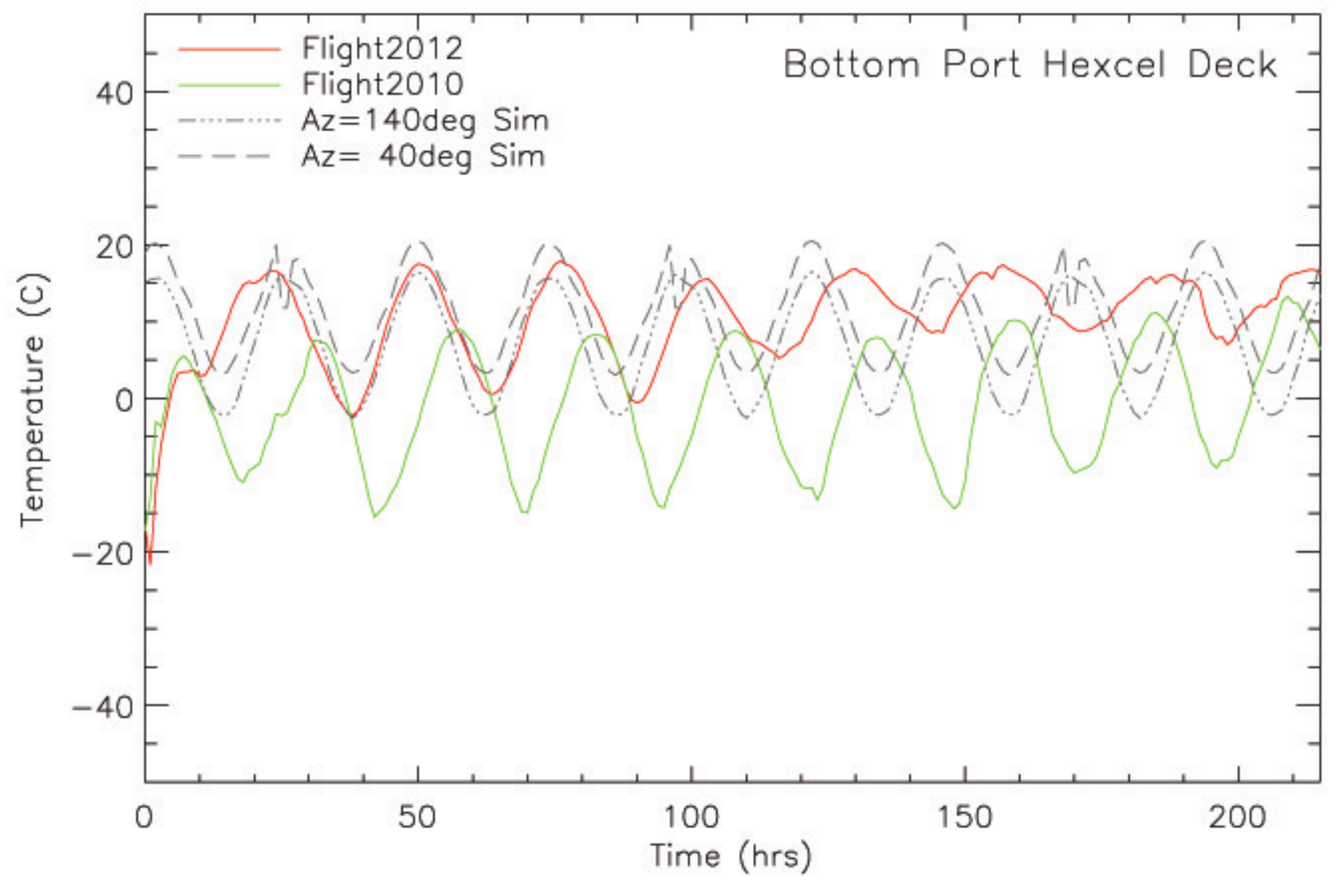}
    \includegraphics[angle=270, width=0.45\linewidth]{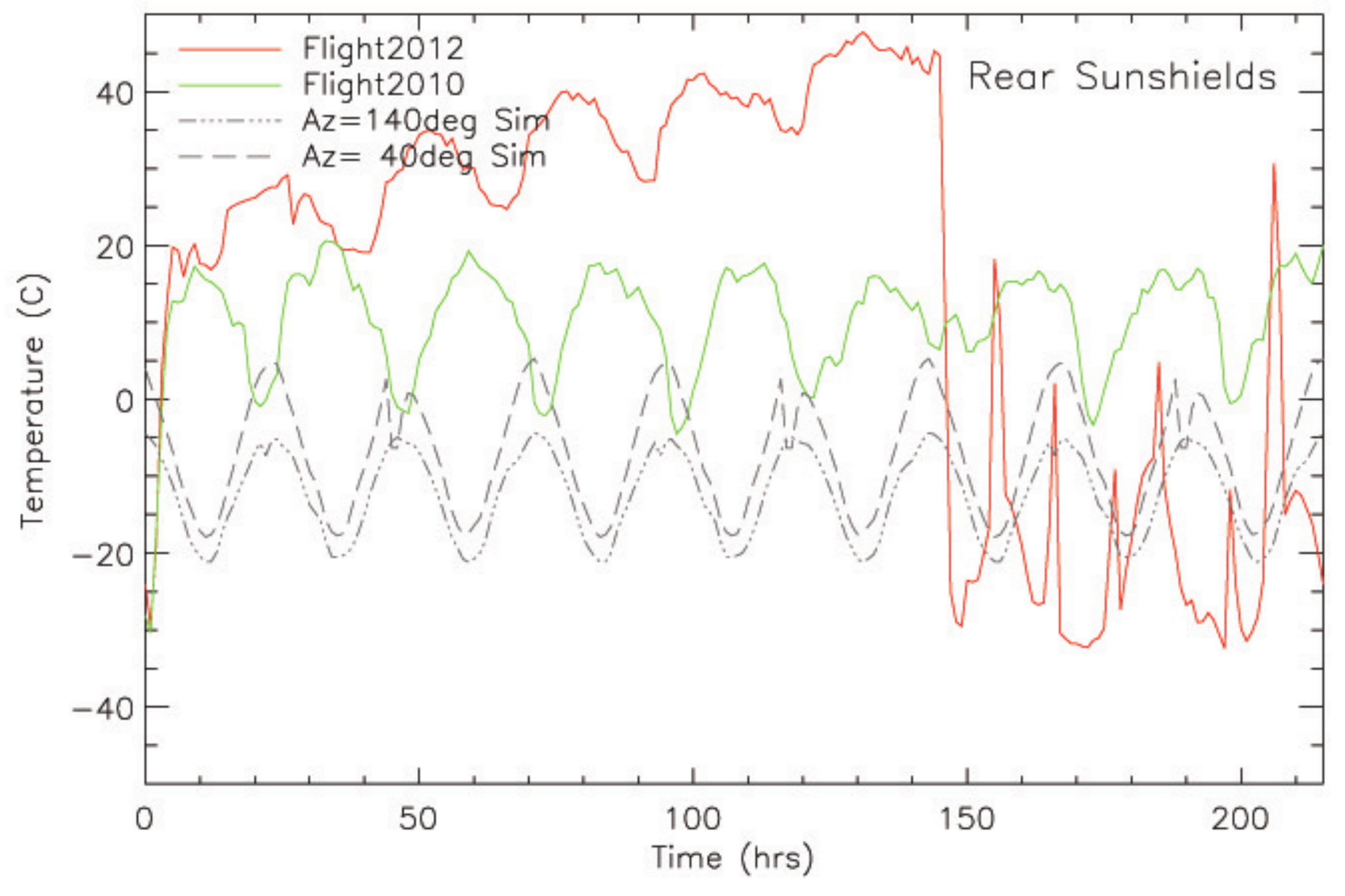}}
 \caption[Flight Temperatures of the BLASTPol Ambient Thermometers]{Flight temperatures of selected passive elements of BLASTPol.}\label{thermal:blastpolambientTplot}
\end{center}
\end{figure}

A series of thermometers were placed in passive elements of the gondola in order to calibrate the thermal model and evaluate the role of convection in the thermal performance of the experiment. This calibration is complicated, but proved successful in elements such as the chin, the port pyramid, and the outer frame, as shown in Figure \ref{thermal:blastpolambientTplot} and summarized in Table \ref{thermal:table1}. More difficult was the estimation of temperature in elements such as the rear sunshields and the solar array, also in Figure \ref{thermal:blastpolambientTplot}, which are boundary layers exposed to convection and are very extended, making its detailed treatment in the model unnecessarily complex and computationally slow.

The solar array is one of the critical elements in the thermal model. Its location is normal to the middle of the azimuth range of the telescope, at 30\degr\ from bore-sight. The solar array is exposed to direct illumination from the Sun on its active face and reflection from the sunshields on the other face. The thermal model in this configuration predicts temperatures of  up to 100\degr C for the surface of the array, in the absence of conduction or convection.

The solar array is simulated by two $0.07\,$inch thick G10 fiberglass plates at $1\,$inch from each other and connected by 1\% 6061 aluminum, simulating the aluminum honeycomb structure of the panel. The surface of the front of the panel is SunCat Solar Cell in the outside and SunCat Laminate Black in the inside. The surface of the back of the panel is SunCat G10/FR4 on both sides. This is the same configuration used by CSBF for the thermal modeling of their solar array. The SunCat solar panels are rated to 100\degr C and this was estimated to be adequate for flight, considering that their surface is a boundary layer, that can be cooled by convection. This assumption was valid, as shown in Figure \ref{thermal:blastpolambientTplot}, where the temperatures peaked at 80\degr C. After BLASTPol10, the back of the upper most panels of the solar array presented minor scorching, most likely cause by the reflection of the sunlight on the sunshield and into the back of the array. The temperatures of the solar array during BLASTPol10/12 are not comparable, since the array configuration changed from 18 panels to 15 panels and the thermometers were placed in different locations.

\section{Conclusions}\label{conclusions}

The thermal design of BLASTPol successfully guaranteed the operation of the experiment during two balloon-borne flights over Antarctica. The unprecedented limits in Sun-avoidance achieved in these campaigns provide a landmark in the design of balloon-borne telescopes. The temperatures registered during the BLASTPol10/12 indicate that the thermal model is effective in determining the operational ranges of gondola elements with heat loads, but a detailed modeling of air conduction is necessary to determine the behavior of passive elements.

\acknowledgments     
The BLAST collaboration acknowledges the support of NASA through grant numbers NNX13AE50G S03 and NNX09AB98G, the Canadian Space Agency (CSA), the Leverhulme Trust through the Research Project Grant F/00 407/BN, Canada's Natural Sciences and Engineering Research Council (NSERC), the Canada Foundation for Innovation, the Ontario Innovation Trust, the Rhode Island Space Grant Consortium, and the National Science Foundation Office of Polar Programs. F. Poidevin thanks the Spanish Ministry of Economy and Competitiveness (MINECO) under the Consolider-Ingenio project CSD2010-00064 (EPI: Exploring the Physics of Inflation) for its support. J.~D. Soler acknowledges the support of the European Research Council under the European Union's Seventh Framework Programme FP7/2007-2013/ERC grant agreement number 267934. J.~D.~Soler also thanks Scott Cannon and Taylor Martin for their valuable advice on Thermal Desktop\reg. We would also like to thank the Columbia Scientific Balloon Facility (CSBF) staff for their continued outstanding work.


\bibliography{InstrumentationRefs,jdslib}{}   
\bibliographystyle{spiebib}   

\end{document}